# Individual context-free online community health indicators fail to identify open source software sustainability

Yo Yehudi*[12], Carole Goble[1], and Caroline Jay[1]

[1] Department of Computer Science, University of Manchester, Oxford Road, M13 9PL, United Kingdom
[2] OLS, Wimblington, United Kingdom

*Corresponding author. E-mail: yochannah.yehudi@postgrad.manchester.ac.uk

## Abstract

The global value of open source software is estimated to be in the billions or trillions worldwide[1], but despite this, it is often under-resourced and subject to high-impact security vulnerabilities and stability failures[2,3]. In order to investigate factors contributing to open source community longevity, we monitored thirty-eight open source projects over the period of a year, focusing primarily, but not exclusively, on open science-related online code-oriented communities. We measured performance indicators, using both subjective and qualitative measures (participant surveys), as well as using computational scripts to retrieve and analyse indicators associated with these projects' online source control codebases. None of the projects were abandoned during this period, and only one project entered a planned shutdown.

Project ages spanned from under one year to over forty years old at the start of the study, and results were highly heterogeneous, showing little commonality across documentation, mean response times for issues and code contributions, and available funding/staffing resources. Whilst source code-based indicators were able to offer some insights into project activity, we observed that similar indicators across *different* projects often had very different meanings when context was taken into account.

We conclude that the individual context-free metrics we studied were *not* sufficient or essential for project longevity and sustainability, and might even become detrimental if used to support high-stakes decision making. When attempting to understand an online open community's longer-term sustainability, we recommend that researchers avoid cross-project quantitative comparisons, and advise instead that they use single-project-level assessments which combine quantitative measures with contextualising qualitative data.

# Introduction

Software is often abandoned or shut down, for a range of different reasons[4–6] for example, a sole maintainer has moved on[7,8] or grant funding has ceased. However, some projects are able to withstand these barriers and may remain active and maintained despite adversity.

Sustainability and longevity is of particular interest to organisations that wish to make a good investment when they choose to support software, whether financially or through in-kind time and effort. Organisations facing this decision might be philanthropic or government grantmakers providing funds to a software project[9,10], corporations supporting development of software infrastructure they depend on, or a software developer team choosing dependencies that their software will rely on[11].

## Open vs closed source (proprietary) computer code and software

Open source computational tooling is often created and used by academics in research and teaching[12]. This study focuses on open source software, as opposed to software tools that are *not* shared publicly, also known as "**closed source**" or "**proprietary**" computer code. The computer code behind proprietary tools remains private, often because the creators of this software intend to commercialise the tool, and expect that people would not otherwise pay for it. The behind-closed-doors nature of proprietary software makes it more challenging to study academically.

Sustainability and health are not necessarily the same thing, but it seems reasonable to assume that a healthier project is more likely to exist for longer, and as such may be used as a proxy for sustainability. This project investigates this assumption.

> **Research Question**:
>
> Can individual online community project health indicators accurately identify project longevity over a 12 month period?

# Literature and background

This study spans domains of open science (many of the projects studied were academic research software), open source software, community life cycles, and sustainability for open source organisations.

## A brief history of open source and free software

Open source code and free-to-use software have existed for decades, but the term "Open source" itself was initially defined in 1998[13]. As an ethos, it overlaps with the "Free Software" movement[14], but in some cases they embrace differing philosophies. Open source generally refers to shared computer code for pragmatic reasons, whilst free software focuses on libertarian freedom to use and share the software without any restriction, often cited as "free as in speech, not free as in beer"[15]. Collectively, these related movements are sometimes referred to as: F/LOSS - Free/Libre Open Source Software; FOSS - Free Open Source Software; or OSS - Open Source Software[16].

The "free as in beer" distinction is a noteworthy one. It is important to separate software which can be used at no cost (the "free beer" comparison) from open source software. Open source software can be commercialised[17], and indeed software that is free or low-cost to the user might be proprietary (examples are Freeware[18] and Shareware[19]). Software without monetary cost might have usage restrictions that do not meet the open source definition - perhaps it is free for academic use, for non-commercial use, or has other specific usage restrictions[20].

The Open Source Initiative maintains both the Open Source Definition, and a list of approved licences that meet this definition[21], effectively exploiting copyright law to enforce the ability for computer code creators to share their work freely[22]. While the official Open Source Definition is long enough that we will not restate it in full, a recent more brief definition from NASA's Transform to Open Science curriculum states:

> "Open Source Software is distributed with its source code without additional cost, that makes it available for use, modification, and distribution with its original rights and permissions."[23]

## Open science as a lens onto open source software

Openly sharing science, research, and scholarship may be done for many reasons: practical (open research is by its nature easier to access than closed work); methodological (transparently shared research is easily peer reviewable), and finally, ideological: science can advance the wellbeing of humanity more effectively if we cooperate, actively involve the affected stakeholders, and build on each other's contributions[24].

### What is open science and when did it start?

Open science is a behaviour that occurs throughout the research lifecycle, and not only a final output[23]. There are varying but overlapping definitions provided by UNESCO[25], NASA[26], NWO[27] (the Netherlands' national funding agency), and many others. For the purpose of our study, we use NASA's definition:

> "Open Science is the principle and practice of making research products and processes available to all, while respecting diverse cultures, maintaining security and privacy, and fostering collaborations, reproducibility, and equity."

At the time of writing this, in 2024, open research practices are being embraced by top-down initiatives, as well as grassroots researchers. Many research journals and funders require the research that they fund to be shared openly, and may provide incentives to researchers to encourage them to share their outputs. The UK's national research funding body, UK Research and Innovation (UKRI), for example, encourages researchers to **preprint** their work (publishing it openly before peer review), and requires peer-reviewed papers to be **Open Access** - that is, available to any reader without cost[28]; the philanthropic biomedical funder the Wellcome Trust provides a free-to-use open access publishing platform for all its funded researchers[29]; and NASA declared 2023 to be "the **year of open science**", dedicating $40 million USD to its "Transform to Open Science" initiative that will train 20,000 scientists over five years[30].

Throughout the research lifecycle, a non-exhaustive list of other aspects of open science include pre-registration (sharing plans for research *before* conducting the research); open protocols; open hardware (affordable, maintainable open designs for scientific machines and tooling); and citizen science (also known as participatory research)[25,30]. We will not go into these in further depth; instead we focus on open source code, the focus of our study, and open data, as it underpins open code.

## Data: As open as possible, as restricted as necessary

Data may be shared openly, restricted-access for ethical reasons, unshared beyond an individual or group of creators, or somewhere in between. There are a plethora of ethical and legal reasons why a computational scientist who generally espouses open sharing principles might not share their data. For example, the RSPB in the UK does not publicly share the breeding site of a rare bird for conservation reasons[31], a healthcare worker should not reveal a patient's personal medical information[32], and the CARE principles for indigenous data provide guidelines for storing and handling data that might affect or exploit first nations[33].

In this study, we deposit as much of our own data openly as we can.

## Open software and code in science and research

The open research software movement is, perhaps less mature than the open source or data movements. The UK's Software Sustainability Institute, founded in 2010, brands itself as the first organisation in the world dedicated to improving software in research[34], and UNESCO defined Open Software as part of open science in 2021[25].

Software may be used to analyse, process, visualise, and/or produce research data, but sharing software openly is, in some cases, easier than sharing data openly. In scenarios where software uses data that is not shared for some ethical or legal reason, a researcher may still share the computer code associated with the data, as a way of **sharing their methods openly**. If needed, a researcher may also provide synthetic data in order to make it possible to run the computer code. Synthetic data are "dummy" data, intentionally generated to resemble the original protected data without revealing genuine protected information.

## Measuring open source projects and communities

Open Source code underpins commerce and research worldwide, but may not have appropriate levels of resource dedicated to it, despite being critical infrastructure. A recent Harvard Business School working paper, which surveyed millions of global companies, estimates the open source supply-side value is around $4.15 billion USD, and estimates that companies "would need to spend 3.5 times more on software than they currently do if OSS did not exist" [1].

Even though so much of the world depends on open source code, there have been high-impact security and stability challenges affecting users of these codebases[35]. Examples of these challenges include the OpenSSL heartbleed vulnerability[2], where an estimated 24%-55% of secure servers worldwide were affected by a flaw that allowed attackers to steal information that would normally have been encrypted. This flaw was severe enough that attacks left no traces, and provided access to business-critical information such as secret keys, usernames, and passwords. Another example is the left-pad incident, when withdrawal of an 11-line JavaScript package from the npm package registry resulted in swathes of the internet unable to update their code, including high-traffic commercial sites[3,36].

Recent initiatives look at ways of measuring open source project health, possibly in order to allow projects to identify and mitigate these kinds of challenges, and/or for potential project adopters (e.g. software developers re-using the code) to understand when stability-threatening challenges are present. The following sections explore literature on: ways to measure OSS projects; definitions of project failure and reasons a project might fail; and OSS lifecycle models.

## Metric and measurement systems

Linåker et al[35] note that the science of measuring open source projects is still emergent, and it is not always clear how to *assess* metrics that may have been identified, or which ones may make sense in the context of a given project. They provide a set of over 30 separate project health measures based on a snowball literature search, with different levels of granularity across project code, security, finance, diversity, stability, popularity and communication.

At the time of designing this study in 2020, the only active indicator framework was the Community Health Analytics Open Source Software (CHAOSS) framework, a Linux Foundation project that released its first metric set in 2019, shortly before the ideation of this study[37,38]. Metrics under this framework assess the individuals who create open source communities, the organisations that back them, project culture and governance, and in-person events as well as remote / online interactions. CHAOSS measures are linked with open source tool sets that implement ways to measure these metrics across GitHub, GitLab, and other open source hosting and collaboration platforms.

We explore indicators from CHAOSS further in the methods, results, and discussion sections.

## Open source sustainability and project survival

### Loss of a sole maintainer (Truck or bus factor)

A known risk condition in open source projects is being stewarded by a single maintainer, who may become burned out or overloaded[39,40]. In the literature, this risk is described as the "truck factor" (TF). This metric is named after the risk assessment question "how many critical project maintainer(s) would have to leave the project (i.e. get hit by a truck) before the project was unable to continue?" Numerous researchers have investigated ways to computationally calculate TF from version control repositories[7].

In the CHAOSS metrics, this measure is called the "bus factor"[38].

### What is "survival", "maintenance", "failure", or "death", in open source project terms?

Definitions of open source activity and failure vary throughout recent literature. Lahtinen defines **project death** as an open repository that no longer receives updates, without a specific time threshold defining how long the lack of updates needs to be before the project is declared dead[8]. One such example might be https://github.com/nvie/gitflow - a popular repository with over 26 thousand stars, but with its last commit 11 years ago - despite the fact that the author is still active on GitHub in other repositories.

Avelino et al defines a project crisis event, when all Truck Factor "TF" developers leave a project, and calls it *Truck Factor Developer Detachment*, or TFDD[7]. They further define project **abandonment** to be when a maintaining TF developer has not committed to a given repository for at least one year, but notes that other papers define a threshold for abandonment to be three months, six months, or one year. This definition also asserts that an abandoned or inactive repository might have ongoing occasional updates by low-volume committers. In this scenario, **survival** events occur when new TF maintainers join and re-activate a repository.

Pashchenko et al note that most "inactivity" thresholds are arbitrary, since all projects have different development strategies and release cycles[41].

Coelho et al created a machine learning model with 80% reliability, used to identify projects that are unmaintained on GitHub. By their definition, **unmaintained** projects may still have sporadic and occasional commits. They also note that being able to predict project failure is useful if it is detected sooner, rather than waiting for longer periods of inactivity to elapse[42].

## Reasons for failure

"Why Modern Open Source Projects Fail[43]" identified nine reasons why open source projects might become inactive, divided into three categories: Environmental, project-specific, and team challenges.

| Environment | Project | Team |
|---|---|---|
| Usurped by competitor<br>Legal problems<br>Acquisition | Obsolete<br>Outdated technologies<br>Low maintainability | Lack of interest<br>Lack of time<br>Conflicts among developers |

Table 1: Categories of open source software failures

**Environmental** challenges touched on factors outside of the software teams' control, such as a competitor taking the market away. This was often a large-scale commercial organisation, such as Apple or Google - releasing core changes to their software that made the entire OSS project irrelevant. **Project** factors included using an out-of-date programming language or having an obsolete dependency (outdated technologies), no longer being required at all (obsolete, such as being intended for a very old operating system), or being hard to maintain for some reason. The final **team** category addresses personal and interpersonal challenges, such as a developer no longer being interested in the project, no longer having time to maintain it, or the team of developers splitting up due to a dispute.

## Some projects do not really die

In 2018, Kula and Robles examine software ecosystems, and note that software tooling that is deeply embedded within an ecosystem is more likely to survive in some form, perhaps due to factors such as organisational inertia to move away from established and embedded systems.

They examine several case studies in The Life and Death of Software Ecosystems, looking at projects that were officially discontinued, but once forked or re-written they continue to live on, ten or fifteen years later[44].

One example of this is CVS, a version control system that was officially discontinued in 2008. At the time of writing, the most recent version of CVSNT, a direct successor of CVS, was in April 2023[45].

## Community culture, etiquette, and social inclusion

Many of the indicators we include in this study are designed to assess project culture and inclusivity. CHAOSS notes that bringing in and retaining a broader number of contributors may be related to project health, particularly with its code of conduct and mentorship indicators[38].

Some open source projects have been reported as having an unwelcoming reputation, or have been affiliated with leaders who admit to having been abusive, or have been accused of harassment. Linus Torvalds, founder of the Linux operating system and Git version control, famously apologised for his abusive culture-setting and resigned from his leadership role in Linux[46]. Richard Stallman, founder of the Free Software Foundation (FSF) and author of the GPL open source licence, resigned positions at MIT and the FSF board, after comments deemed inappropriate around the Epstein child sex scandal[47].

Whilst these are particularly extreme examples of unwelcoming or toxic culture, studies indicate that open source communities can knowingly or unknowingly drive participants away[48–51]. An example of unwelcoming behaviours includes open source contributors who do not get their initial contributions accepted into the repository, and feel sufficiently disheartened that they do not return or try again.

**Destructive criticism** is defined as feedback that is not "specific, considerate, and [does] not attribute poor performance to internal causes"[52]. This type of criticism can drive individuals to leave a community, and is more likely to drive women out of an online software development community than men[53], suggesting there may be other groups that are systematically excluded by cultural norms. This is consistent with observations from a 2017 survey of GitHub users, who were 90% men, and only 15% identified as a minority in the country they lived in[54].

## Software lifecycle frameworks

Lehmans' Laws of Software Evolution divides software into categories, noting that most programs are intended for sufficiently complex tasks that they must continually evolve in order to remain satisfactory, based on iterative feedback and desired features. Lehman defines these evolving programs as "E programs", and programs that can be fully defined by an initial specification as "S programs"[55].

Avelino et al note that most programs fall into the E program category, observing that occasionally mature software systems can be deemed feature-complete, but may still need bug fixes and security updates[7].

Champion and Hill define an open source risk of **underproduction**: the scenario where a package has an imbalance between its need for quality improvement, and the labour of people to implement these improvements[56].

Wynn looks at established business literature focusing on organisational structure, hierarchy, and sales growth/loss over time, conjecturing that downloads over time might be an appropriate measurement for open source lifecycles[57]. Wynn's model is formed of four stages: **Introduction, Growth, Maturity, and Decline/Revival**. Projects may move between Growth, Maturity and Decline/Revival repeatedly, depending on organisational circumstances.

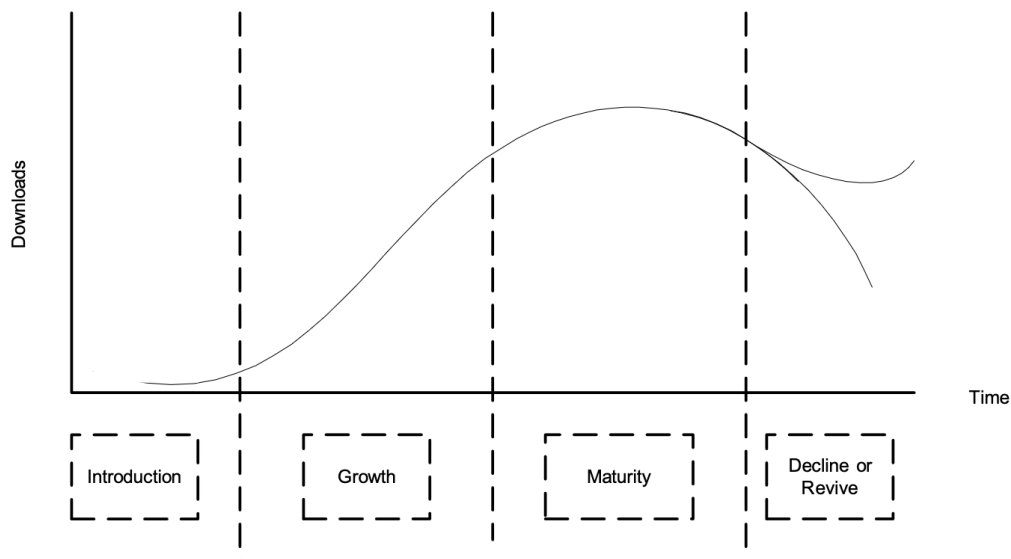

Wiggins and Crowston[58] introduce a model similar to Wynn's Introduction and Growth stages, and add three classes of outcome to each stage, resulting in projects that might be any of the following tragedy / growth / indeterminate states:

|  | Phase: Initiation | Phase: Growth |
|---|---|---|
| Outcome: Indeterminate | Indeterminate, Initiation (II) | Indeterminate, Growth (GI) |
| Outcome: Tragedy | Tragedy, Initiation (TI) | Tragedy, Growth (TG) |
| Outcome: Success | Success, Initiation (SI) | Success, Growth (SG) |

Table 2: Tragedy/success lifecycle framework

We explore applications of Wynn's four-stage model and Wiggins and Crowston's tragedy/success model in the discussion of this chapter.

# Methods

## Ethical review

This study was approved by the University of Manchester Department of Computer Science ethics panel, approval review reference 2020-9098-15844, amendment references 2021-9098-21098 and 2022-9098-25426.

## Recruitment

Participants were recruited primarily through social media calls for participation, as well as Slack groups and mailing lists, targeted towards open source software engineers, community members, and project leaders. No specific domains or programming languages were targeted, beyond a call for "open source and open science projects".

## Study Design

This study used both participant-reported data (surveys), and project activity indicators from project version control repositories that were publicly accessible on GitHub.

## Surveys

We used Qualtrics[59] to administer all surveys. Surveys were administered at the start ("Month 0"), midpoint ("Month 6"), and end ("Month 12") of an approximately 12-month period, from mid-2021 to mid-2022. Response periods for each survey were open for six weeks.

The month 0 survey contained two parts: First, an initial participant information sheet, consent form, and basic information about the repository, such as source code repository user id, URL and project age - elements that were relatively unlikely to change over time.

The second part of the survey gathered temporally sensitive information about the respondents themselves, their projects, user numbers, funding and staffing levels, and expectations for the future. This section of the survey was re-administered in months 6 and 12.

This data contextualises information gathered in the activity indicators portion of the study. For example, a project that catered to a core community of 200 geophysicists worldwide is unlikely to desire or expect the same level of activity as a well-used open source library such as pandas[60,61] or ggplot[62].

The full set of survey questions is listed in the appendix. Pseudonymised responses are deposited on Zenodo, <u>excluding</u> responses from participants who declined the optional prompt in their consent form: "You can still participate in the study even if you choose to answer 'no' to the next item: I agree to share my pseudonymised survey responses openly online. (Optional)."

## Activity indicators

When selecting indicators to collect from the provided codebase URLs, we focused on community engagement indicators (the people who create the open source code) and codebase activity indicators, drawn from version 1 (2019-08) of the CHAOSS metrics, as well as consultation with domain experts within research software engineering and open source.

| Indicator | Source | Indication of sustainability | How assessed |
|---|---|---|---|
| Licence | CHAOSS<br><br>Expert suggestion | Licences create a legal way for others to reuse the code / creative work in question.<br>Domain expert from the Software Freedom Conservancy suggested investigating whether copyleft vs. permissive licences made a difference in results | Source control: Presence of a licence file, or a statement referencing an external licence |
| Contributing guidelines | Expert suggestion<br><br>GitHub project health measure | GitHub's "Community" guidelines https://docs.github.com/en/communit ies/setting-up-your-project-for-health y-contributions/setting-guidelines-for -repository-contributors | Source control: Presence of "how to contribute" guidelines |
| Mentorship | CHAOSS | Actively bringing in new maintainers and contributors seems likely to reduce the "Truck Factor" | Survey: Internship questions. |
| Code of Conduct for Project | CHAOSS, GitHub project health | Community culture re attracting and retaining contributors | Source control: "Code of conduct" or "participation guidelines" |
| Number of commits | CHAOSS | A change in pattern could indicate lifecycle events, e.g. growth, a new approach, or a crisis | Source control, survey. |
| Number of pull requests | CHAOSS | | |
| Number of committers | CHAOSS | | |
| Time to close - PR/Issue | CHAOSS + Expert suggestion | | |
| Grant funding | Expert suggestion | Resources required to build software | Survey: Grant funding questions |
| Staffing | Expert suggestion | Resources required to build software | Survey: Paid staff question |

Table 3: Selected activity indicators

The indicators we did not select from this version of the CHAOSS metrics were either impractical to measure, or extremely industry-oriented, and unlikely to apply to most of our studied repositories' use cases.

Industry-oriented measures we did not select:
- **Elephant factor** - the number of companies that contribute 50% or more effort. Challenging to assess and unlikely to apply to smaller packages and/or research software.
- **Licence file coverage** - how many individual files in the codebase have a licence statement. In the CHAOSS documentation, this was described as a possible measure of business risk that might prevent a company from adopting a piece of software.

Impractical:
- **Testing** - test suites are implemented in many different ways, and challenging to quantify.

## Sample selection

The "Month 0" survey used an initial filter: respondents were screened out via the screening question shown below. Respondents who passed this screening were assumed to answer truthfully and with authority about their project's current status and known plans for the future.

**Screener question:**

*(Only respondents who answered "yes" to this question were asked to read the participant information sheet, sign the consent form, and asked any follow-up questions)*

Are you leading or co-leading an open source / open science related project?

This might mean you're a research software engineer who writes open source software, a leader of an open science community project or event, or a researcher who writes code in R, Python, or any other coding language, so long as the project is open. You don't have to be the PI of the project, but it's best if any given project only answers once.

> If you're unsure if this applies to you, please contact [first author] for clarification

**Terminology:** The term "participant" could easily refer to either the project leader who responded to the survey, or to the open source projects that they lead. In order to avoid ambiguity we will refer to "respondents" and "project leaders" when referring to the individual who responded to the survey, and "participating project" or simply "project", when referring to the project that they lead.

For each of the follow up surveys in months 6 and month 12, respondents were emailed once by the Qualtrics email survey distribution system, and in the case of non-responders, one additional personalised email asking them to respond. None of the non-responders from month 6 failed to respond in month 12, meaning we have at least two of the three survey responses from all respondents.

## Data Preparation and Analysis

### Survey data - cleaning and anonymisation

Survey data underwent three preparation steps using OpenRefine[63]: one for data tidying and two data anonymisation steps. We share both the data itself and the metadata history which describes the tidying steps, to make the methods as transparent and reproducible as possible.

The second anonymisation step was necessary due to the way OpenRefine records data transforms. It is designed to allow a researcher to record and replay most data manipulations by recording detailed steps of all processes applied to the data. For example, if a tidying/first-pass anonymisation step converted "University of Manchester" to "[Anonymous Institution]", this meant that the primary study data was now anonymous, but the metadata describing the processing steps would still have preserved the phrase "University of Manchester".

### Activity indicators data

Survey participants were asked to share one or more URLs to version control for their project. Most respondents provided one or more GitHub URLs. Only two projects provided URLs for other source control servers, such as GitLab, and Sourceforge. See *Results > Respondent characteristics and expectations > Source code repositories* for more detail on provided URL types.

Given the large majority of projects were from the same provider (GitHub), we scripted access to these indicators computationally. We <u>did not gather activity indicators from non-GitHub projects</u>, but <u>did include non-GitHub projects in the aggregate reports on survey data sections,</u> as we still had full survey responses.

After each of the survey periods had closed, we ran scripts to gather our chosen indicators over each GitHub URL provided. This was necessary due to the fact that not all GitHub API methods allow specification of a time period, e.g. the License API[64], which at the time of writing returns a current licence, and offers no option to modify queries by date.

In cases where GitHub's API did allow temporal filtering, such as the Commits, Pull Requests, and Issues APIS, we gathered some results after the study period, time-fenced to the period of the study itself. Instructions to run this computer code are clearly documented in the readme file for the computational analysis code repository.

Activity-indicator gathering scripts all used a GitHub-provided Javascript API library, octokit.js[65].

## Verification: indicator scripts and data results.

All activity indicator scripts that process data include computational unit tests to verify the methods produce expected results in a consistent way. Unit tests run automatically (continuous integration) when computer code updates are pushed to their source code repository.

Note that these scripts test only internal processing of data, after it has been downloaded from the GitHub API - for example, a unit test might verify that a mean average value is calculated correctly, but does not test the results of the API themselves. Verification that the data in the API returns correct results was performed manually in the next step.

Whilst GitHub's API is an effective way to systematically bulk-gather data, not all indicators are computationally gleanable. For example, GitHub can easily report on licence files found in standardised files, such as COPYING, LICENSE, or licence.md, but if a project provides a licence in a non-standard location, such as docs/licence.md, or using a non-standard licence, GitHub might report that the project does not have a licence when it actually does.

Positive results were usually easily verifiable. GitHub's API returns links to individual files or commit hashes when it finds them. We illustrate two examples of this in the following table:

| Measure type | Quickly verifiable "positive" results from the GitHub API | "Negative" not found results - **requiring manual verification** |
|---|---|---|
| Licence | `https://api.github.com/licenses/apache-2.0` | `NOASSERTION` |
| Commit | `https://api.github.com/repos/octocat/Hello-World/git/commits/6dcb09b5b57875f334f61aebed695e2e4193db5e"` JSON purposefully truncated, as the result is long | No commit hash links returned if commits are not found. |

Table 4: Example of easily verifiable positive results and potential false negative results from the GitHub API

When verifying results, we spot-checked the results returned by the API, particularly the negative results, to check for false negatives. False negatives returned by the API are reported in the results section, particularly the following sections: *Source control: Licences*, *Source control: contributing guidelines*, and *Project culture: behavioural expectations in codes of conduct.*

When manual checks like this were necessary, we exported the data to OpenRefine to review, tidy, and update the data. We share the metadata preparation steps for this phase on Zenodo, but not the data itself, due to the need for privacy preservation.

## Results tables and data visualisations

This study focuses on open source and open science projects. In order to make results for this study as open as possible, we created website-based data visualisations and tables, as well as making the analysis code open source.

The project website was generated statically via Jekyll[66], which manages some of the data filtering and presentation. Due to the fact that Jekyll's templating language, Liquid, does not offer the ability to create complex logical statements, the majority of data processing (calculating aggregate results for tables, and data formatting for graphs) was completed using Javascript. Graphs were created using the open source charting library Chart.js[67].

Code for the activity indicator data drawn from GitHub is shared (see data and code availability section below) even when the visualised data itself is not shared.

## Data and Code Availability

**Survey data** are available on Zenodo, DOI https://doi.org/10.5281/zenodo.7347763, licensed CC-BY 4.0 - most participants agreed to share pseudonymised survey data when they signed up to the study.

**GitHub activity indicator** data are <u>not</u> shared openly in order to protect respondent privacy and comply with ethical guidelines. Where anonymised data-processing metadata from OpenRefine is available, we bundle it with the Survey data above on Zenodo.

**Computational analysis code** is deposited on Zenodo at https://doi.org/10.5281/zenodo.8082468 and on GitHub at https://github.com/Sustainable-Open-Science-and-Software/sustainable-communities-tracker. Computer code is shared under the MIT licence.

All resources above, including the **visualisation website**, are linked to from https://sustainable-open-science-and-software.github.io/

# Results

## Meta-results: observations about the study itself

### Privacy, consent, and ethics in open source

This study was launched in the wake of an open source research scandal, when the entirety of The University of Minnesota was banned from contributing to the Linux Kernel, after a group of researchers repeatedly made attempts to get malicious computer code introduced into the kernel codebase. Their intent was to use these commits as part of their research study into security practices of open source computer code, but the researchers persisted in attempting to get their intentionally bug-laden computer code merged, even after maintainers explicitly asked them to stop[68].

While this event happened after our study had already obtained ethical approval, it emphasises the importance of dealing with the community in good faith. We planned our study with the intent of being transparent and clear about our plans to study these projects - something that can be a challenge when an open source repository might have tens, hundreds or even more contributors. The largest of projects participating in our study had more than 1200 individual contributors, over the history of its source control.

Our approach to provide a reasonable level of transparency and consent involves three major components: Consent from a project leader, a prominent notice on the repository about the study, and sharing indicators drawn from GitHub at an aggregate project level, *not at an individual level.*

At the end of the first survey, participants were prompted to add a research notice to their repository's readme (Available in *Appendix > Research notice*). Some project leaders did so, but many did not notice the request, or did not follow up on it. Before running any of our indicator-gathering scripts, we systematically checked the repositories for this notice, and if missing, added it to the repository as a pull request. In order to make sure this did not skew our data, we made sure to manually remove that pull request/commit from reported activity metrics.

"I think you're a bit overcautious here" -- anonymous survey respondent

Many participants merged the pull request without any further comments or challenges. Some felt that the readme was not the correct location, or that we had chosen the wrong primary repository, if multiple repository URLs were provided. In these cases, we were able to negotiate an acceptable location, such as an issue on the repository, supplemented by emailing the project's mailing list.

Not all participants, however, were happy being asked to add this notice. Some (as quoted above) felt that while the intent was good, the level of caution was higher than necessary, citing the fact that their project was already open. In other cases, it was necessary to appeal to the project leader who originally filled in the survey in order to get the notice added, as other maintainers were vehemently against adding a notice in their repository.

This meta-level data was not gathered systematically, as it was not a challenge we foresaw, but we would recommend that others take this kind of challenge into account when planning similar studies.

## Not all GitHub API results are of equal quality

Misleading results returned by the GitHub API are a known challenge[69]. In our case, this was largely represented by the false negatives, when open source community "good standards" such as licences, contributing guidelines, or codes of conduct were not found by the API, but could easily be located by a human looking at the repository. In cases where false negatives were present, we spot-checked the positives, and manually reviewed all negatives, correcting false negatives, before interpreting the data.

We also observed that the community endpoint[70] results changed over time. This endpoint is an analogue of the community page in GitHub's "insights" section, that prompts repository maintainers to add specific policies and template files that it recommends for community health.

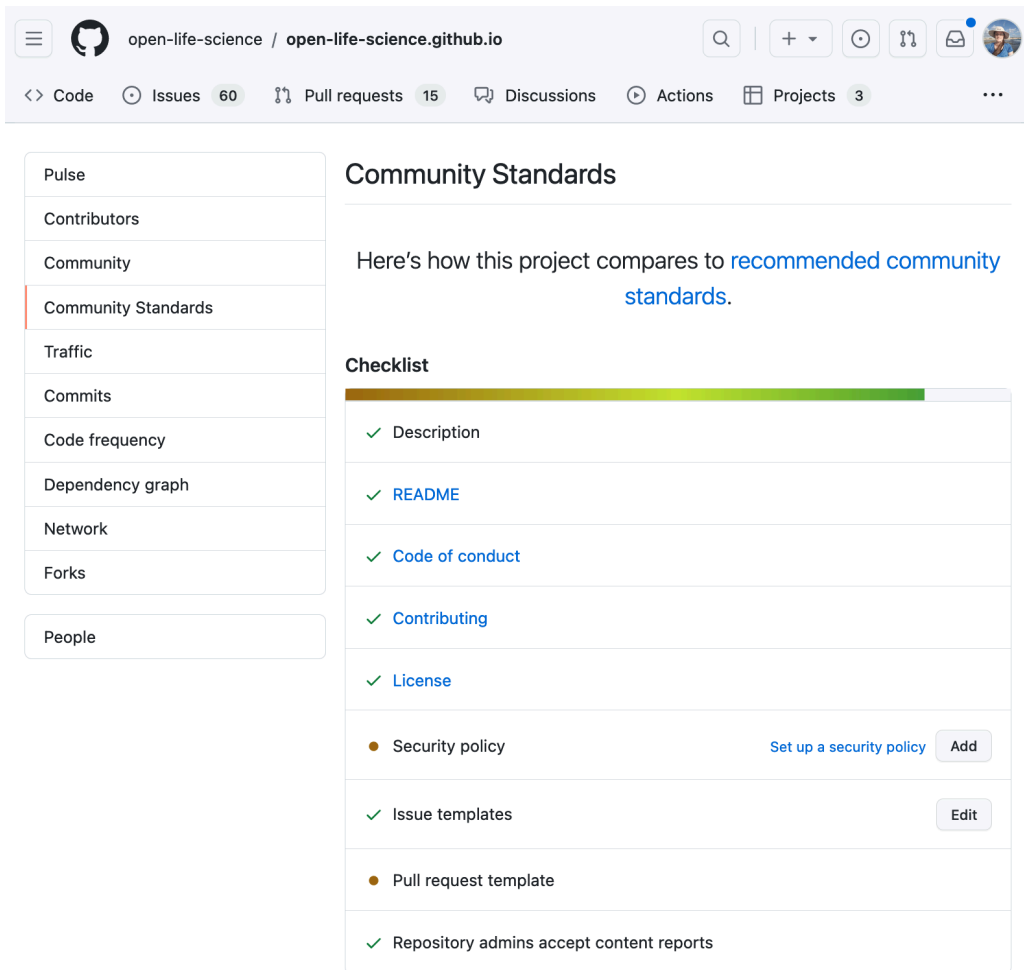

Figure 2: Community standards GitHub page, as shown for the repository at https://github.com/open-life-science/open-life-science.github.io/community

In our observations, some repositories that had a score of 100% at month 0 had their score go down over time, despite *no change* in the presence of files such as code of conduct, readme, and contributing guide. We did not use the community health percentage in this study due to this variability.

## Response rates

In total, the original "Month 0" survey received 140 responses. Over two thirds of these (102 respondents) did not complete the full survey: they were screened out via the initial screening question, dropped out of the survey before completing the consent form, or did not provide a URL for their project. We call the remaining 38 responses "participating projects".

**Response rates for follow-up surveys in months 6 and 12**: All respondents responded to the initial survey and at least one of the follow-up surveys. Most participating project leaders responded to both of the follow-up surveys. The exact breakdown is shown in the following table:

| | M0 | M06 | M12 | Respondents who responded to all three surveys |
|---|---|---|---|---|
| Initial Survey responses | 140 | - | - | - |
| Responses that passed the initial screening question | 115 | - | - | - |
| Completed responses *(exact number)* | 38 | 34 | 34 | 29 |
| Complete subsequent responses *(as a percentage of initial complete responses)* | 100% | 89% | 89% | 76% |

Table 5: Number of responses to the survey components of this study.

## Participant profile: Project age

> **Survey question:**
>
> When was this project founded? It's okay to approximate, we don't need a precise day.

The respondent base had a strong skew towards newer projects. Of the 38 participating projects, most projects self-reported as having been founded in the preceding twenty years, with only two projects reporting their founding date as prior to 2000: the oldest projects reported they were founded in around 1975 and 1998. The mean project age was approximately 8.5 years, and the median was 5.4 years.

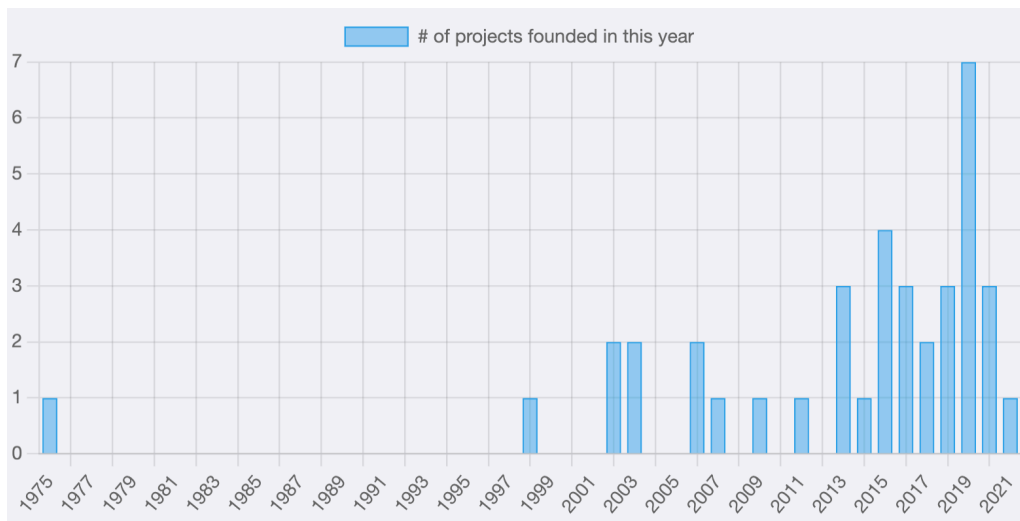

Figure 3: Histogram of project founding year.

| | |
|---|---|
| **0-4 years** | 16 |
| **5-9 years** | 11 |
| **10-19 years** | 9 |
| **20+ years** | 2 |

Table 6: Project age separated into five-year buckets

## Participant profile: Project domain and project type

Recruitment did not discriminate based on domain, beyond calling for open source or open science projects. As such, while most participating projects were scientific software or computer code of some kind, the exact scientific domains were varied, and included mathematics, ecology and environmental science, bioinformatics, and neuroscience.

Some of the projects were cross-domain tools commonly used in research and science, such as HPC libraries, computational notebooks and popular libraries used for statistics or data processing.

A small subset of projects were open source projects likely to attract primarily (but not exclusively) non-researcher users, such as open source operating systems or cluster management.

While we report briefly on these project domain details, they were not gathered systematically, and we do not segment the data based on these characteristics.

## Participant profile: Source code repositories

**Survey question:**

What is your project's online code repository URL (e.g. GitHub URL). If you have more than one repository for your project please share as many URLs as you think are relevant, one per line.

Across 38 projects, 37 provided a GitHub URL, with the remaining single project providing a GitLab link. Most projects provided one single repository link, e.g. `github.com/manchester/my-project`, and a small proportion provided multiple repository links, GitHub organisations without a specific repository link, GitHub and Sourceforge links, or project websites that were not for a version control repository.

A breakdown of URL types is shown in the following table. Note that the total adds up to more than 38 as some projects provided multiple types of URL, e.g. two GitHub URLs and a Sourceforge URL.

| URLs provided | Quantity | Notes |
|---|---|---|
| Single GitHub repository | 28 | |
| Multiple Github repositories | 8 | |
| GitHub org without repository | 4 | |
| Sourceforge repository | 2 | |
| GitLab repository | 1 | |
| Non-version-control site | 3 | Docs site, general application website, or package repository such as pypi or CRAN |

**Table 7**: Number and types of URLs provided by project leaders.

Note that because some repositories provided multiple source code URLs, the number of results for version-control indicators is often higher than the number of results for the survey at a given time point.

## Activity indicators

The bulk of this study's activity indicators data are temporal - that is, data which changes across the three timepoints of month 0, month 6, and month 12. Visualisations of these data over time, individual case studies, points of interest, and aggregate summaries of the raw data are presented in the "Data Annexe" as part of this paper's appendix. Interpretations of the data are presented in the following results sub-sections: Survey theme, Meta-results, and Interpretation and answer to the research question.

## Survey theme: Who is "paid" to work on an open source project?

An open source project may be open to collaboration from people affiliated with a range of different organisations, or indeed who were working independently without any organisational backing at all. This leads to a recurrent theme from respondents, who were unsure how to define who was paid on the project. Many contributors or project leaders used in-kind time to work on the project in the course of their job, but were not officially paid or endorsed to work on the project. Others were "permitted" to work on the project by a supervisor or manager using work time, but were not paid using funds allocated to the project. Illustrative comments from across all three survey timepoints are shown below:

> "Unsure about this, though the tools in this project had been used extensively (in one form or another) during the early days of genome sequencing and analysis, including the Human Genome Project, so there is very likely code that benefitted from grant funding." - Project 15

> "Nobody was directly paid, but updates were made as part of their job as the research institute" - Project 28

"Several of the questions asked about paid work for contributions to the project; I answered 0 because nobody besides myself has been explicitly paid to work on it, but many have contributed to it under the broad umbrella of their relevant project work, for which they are being paid." - Project 31

"1-3 (1 full time, several people are able to spend work hours on it) - Project 73

"Contributions to open source count towards my job [as] a departmental lecturer" - Project 79

"My academic employer accepts that I spend time on this project" - Project 106

## Interpretation and answer to the research question

Research Question recap: "Can individual version-control project health indicators accurately identify project sustainability over a 12 month period?"

All but one of the projects was still open to contribution at the end of the study. No projects appeared to *fail* completely (that is, no projects shut down due to lack of resources), and the sole project that did close down was a planned sunset that matched with the respondent's original plans from month 0. While we were therefore unable to meaningfully compare *failed* projects with sustained projects, we were still able to examine indicators for older and/or larger projects (which have proven sustainability), and compare their indicators with younger projects. Literature supports the assumption that some of our younger studied projects are likely to fail[4,42–44,71–74] before reaching the age of the older projects. Given this expectation, the variance between older projects in the study, variance in development practices, and the similarities between indicators from older and newer projects, **we conclude that single indicators are unable to predict project sustainability and can not be meaningfully compared between unrelated projects.** Table 8 below summarises the indicators examined and provides pointers to supporting evidence from the results section.

| Indicator | Assessment method(s) | Essential for sustainability? | Evidence |
|---|---|---|---|
| Licence | Source control: Presence of a licence file, or a statement referencing an external licence | No | Projects may have multiple source-control repositories, some of which may be lacking licences. This indicator needs to be assessed holistically across the whole project ecosystem. *See Table 15.* |
| Contributing guidelines | Source control: Presence of "how to contribute" guidelines | No | Roughly two-thirds of projects had some kind of contributor guidelines; this was evenly distributed amongst the older projects which had sustained for longer, as well as across new projects. *See Open to contribution: Source control: Contributing guidelines.* |
| Mentorship | Survey: Internship questions | *n/a* | Few of the projects studied had interns - insufficient information to judge. *See Ffigure 7.* |
| Code of Conduct for Project | Source control: "Code of conduct" or "participation guidelines" | No | Roughly two-thirds of projects had some kind of code of conduct (CoC) guidelines; this was evenly distributed amongst the older projects which had sustained for longer, as well as across new projects. The two projects which added CoCs during the study were 15+ years old, so their historic sustainability can not be attributed to the CoC. *See Ffigure 8.* |
| Number of commits | Source control & Survey | No | Low and high numbers of commits were found amongst both older (and therefore longer-sustained) projects, and newer projects. *Examples: Projects 34, 16, and 113 all had around 300 commits during the period of the study, but were respectively founded across a nearly 20-year period (2019, 2015, and 2002). See Figure 11.* |
| Number of committers | | | *See staffing below.* |
| Time to close - PR/Issue | | No | Mean times to close varied amongst both older, established projects and newer projects, with many similarities between older and newer projects. Literature suggests this is likely related to project-specific workflow behaviours. *See figure 10.* |
| Direct grant | Survey: Grant | No | Self-reported answers were varied. Some of the oldest |
| funding | funding questions | | projects had no directly awarded funding for the entire study.<br>*See* Figure 14. |
| Staffing | Survey: Paid staff & leadership questions | No | Oldest and newest projects had similar leadership team sizes.<br>*See* Figure 6, Table 20, Table 21. |

Table 8: Results concluding summary table

# Discussion

| Sustainability indicators in an open online community | |
|---|---|
| **Research Question** | **Short answer** |
| Given contextualising information about open source project funding and staffing, can project health indicators accurately predict sustainability over a 12 month period? | No. In our study, we did not find evidence that indicators alone can predict project health. Projects are heterogeneous and have a range of variation in resources available, goals, desires, and user bases.<br><br>This context, along with project age, culture, and development methodologies, means that while project events may be *detectable* using these indicators, changes in health indicators may have different explanations, which may or may not be gleaned from source control alone. |

Table 9: Research question recap

We review three primary themes in our discussion: heterogeneity; misaligned expectations and resources; and context.

Open Source is not a monolith, but instead a loose collection of projects with a similar philosophy of sharing their code openly and freely (and that is free as in "free speech", not "free beer"). They have **heterogeneous** goals, behaviours and contributors. In the following sections we examine: the concern from participants that they did not conform with open source norms (and possible resultant self-selection bias); the heterogeneity of changing standards over time; and the varying definitions of survival and failure.

In addition, the desires, expectations, and resources that our project leaders wished for were often **misaligned** with what was actually available. Few project leaders actually expected to close down their projects if they were no longer able to maintain them, and top-down incentives are poor: projects might be used by hundreds of thousands with very little financial support.

Finally, source control is often **context-poor**. When we examined the source control, we were able to get part of the picture with regards to the activity a project carries out. By running surveys with additional questions that we couldn't access via source control, such as funding sources and paid staff, we were able to glean more of the picture about each repository, and what its sustainability trajectory might be. Knowing the age and history of a project made it easier to interpret metrics drawn from source control - but *without that context*, it was almost impossible to draw meaningful insights from the source control indicators.

## Heterogeneity: "Our repository is different - should we participate?"

More than one participant asked this type of question before - or even after - signing up to participate in this study. These participants worried that using a source control repository in what they perceived as a non-standard way would mean that their repository should not be included in the study. For example, in one case, the participant reported that materials were developed jointly but only committed by a single individual. In another, the primary location where collaboration happened was a private server, which was then mirrored to a public repository.

In every case, the participants who asked this question were running open science and/or open source projects, so our answer was always "Yes, please do - your results will probably be interesting." It seems likely that others in the same situation may have self-selected out of participation without asking first. Rather than limiting our study to projects that did open source "right" in some way, we chose to investigate the individual projects and attempt to understand these outliers, to glean some of the context and nuance that creates these different project types.

## Heterogeneity: When were the 'standards' written, and by whom?

Whilst GitHub is nearly ubiquitous as a hosting and social platform for code in our current (2021-2022) study, the version control system Git was developed in 2005[75], and GitHub in 2007[76]. Six of the study repositories were founded before Git was created, and nine in total were founded before or during the year GitHub was launched. It also seems likely that some of the projects founded later than 2007 used a different platform, before the decline of Sourceforge[77] and GitHub gaining traction.

GitHub is by no means the *de jure* standards body for open source convention, if indeed there is one. But given the ubiquity of open source projects hosted on GitHub, it seems likely that GitHub is a current *de facto* standards body. If a new project is guided to add a licence, a readme, and other useful introductory files, this may indeed benefit the projects and be helpful for future contributors. Projects that have been around since *before the platform existed* exemplify genuine longevity and arguably may demonstrate good practice, even if it does not align to current norms.

Nevertheless, older projects may struggle to be defined as "healthy" if these indicators are relied upon, without looking deeper into the reasons for the differences. A project founded in 1997 might have a non-standard (but open) usage licence, from before the Open Source Initiative was founded and began to take steps to reduce licence proliferation.

A project that lacks a CONTRIBUTING.md file might be a brand new project, written by someone who is not yet familiar with current conventions. Project 85 might be an example of this, at less than a year old when the study was conducted. But equally, it might be a mature project like Project 15, that is over 20 years old, has a large number of committers (70 recorded in Git/GitHub, and probably others before the project migrated to Git), and has heavily documented code that expressly invites contribution.

## Heterogeneity: Open Source Project "Survival", "Maintenance", "Failure" and "Death", revisited

In the literature section, we reviewed some of the many ways these terms have been used in previous studies. One thing is very clear both from the literature and from results from this study: different projects have different needs and expectations.

The activity on small, newer projects may often be hard to distinguish from mature projects that are slowing down. Some older projects were bustling with activity, whilst others had occasional updates. They may also have similar leadership profiles, with only one or two leaders - but for very different reasons. In a newer project, the founder(s) often have yet to establish governance and sustainability plans, and may only do so if the project grows and formalised governance is necessary. In an older project, there may have been larger leadership teams in the past, but if the project is in low-activity maintenance mode, having a large board stewarding an alive-but-slow project would be a poor use of limited resources.

## Misalignment: Intent vs reality vs need

We asked participants what their projects might be like in five year's time, if resources weren't an issue (see figure 4 for a full list of responses). In these best-case scenarios, our respondents usually reported that they saw the project continuing for at least five years, whether they themselves were part of the project or not. This seems like a reasonable and realistic hope, given that they were asked to conjecture on a scenario where they had all the resources they needed.

In worst-case scenarios (Survey question: "If you find yourself unable to maintain the project any longer, what do you foresee happening?", table 10), however, participants were generally very optimistic. Around half of participants thought their community would keep the project going, half thought that they would be able to contribute in some way in their own free time, and only one or two respondents thought that they would close the project down.

In reality, many open source projects fail or are abandoned by their original maintainers[4,42–44,71–74], in some cases without any notification that the project is being discontinued. This is why some "abandonment" metrics require up to a year to elapse with no updates before defining a project being no longer maintained[42].

The seeming optimism of our participants could be down to self-selection. Perhaps only people who wished to see their project survive long term applied to participate in a project that studied sustainability. But equally, it seems possible that many people overestimate their capacity to maintain a project without the backing of dedicated resources or funding.

This also raises another question. Just because a project exists now, does it need to continue to exist indefinitely? Very few of the participating projects viewed their work as finite works with a clear end point. It is hard to measure project utility, so we may substitute proxy metrics such as number of users or number of citations instead. But the open source world is not meritocratic - it is documented to have charismatic leaders whose fame results in quick adoption of projects they create or endorse.[78]

Philanthropic program officers from the Sloan Foundation and the Chan Zuckerberg Initiative note that in a setting with limited financial resources, inevitably some software projects will have to be discontinued[9]. The "sunk cost fallacy" suggests that having invested in something in the past is insufficient justification for investing in it in the future. Equally, it seems likely that resources are inefficiently spent if we do not maintain high-quality, mature, well-used software tools. Scientists value novelty, but a software package is only novel once, and inherently ceases being novel as time passes - even though it may remain *valuable* within the scientific ecosystem.

## Misalignment: Sustainability vs resourcing in open source research projects

Whilst open source may have originally been designed as a pragmatic way of exploiting free labour[14,79], a common but incorrect assumption is that open source projects do not earn money and community members are never paid to work on them[80,80,81], or that because they are volunteers, access to resources and long hours with little pay is not an issue[43].

Responses from project leaders at the end of the study showed that around 41% of respondents *had* active grant funding, and 62% *wished to have* more funding - directly contradicting assumptions that open source work is always provided for free.

The majority of respondents reported a desire for more resources, and even the most-used project - Project 73 - had users numbering in the tens of thousands or millions, but only had partial salary support for its developers. Given this mis-match, it seems plausible that one strong reason for open source project "failure" is due to under-resourcing, resulting in burnout of maintainers - a serious and studied issue in both open source and in academia[82].

The Amsterdam Declaration on Funding Research Software Sustainability (ADORE)[71] notes that funders of research software generally fund on a temporary short-term basis, and do not necessarily recognise the need for a project to be maintained if it is to be of ongoing use after the initial funding period. Indeed, funders have historically noted that projects are often more likely to be invested in if they are "novel"[9], which seems to be in direct contradiction with the need for scientific software to be robust, updated, and reliable, if it is to produce correct scientific results.

## Danger Zone: Inferring meaning from version control and social network-based indicators requires in-depth context

As scientists researching open source behaviours, the desire to use computational scripts to easily assess software repositories is appealing. Software engineers are used to writing code to resolve their challenges, and scientists are regularly measured by the quantity of their publications, or the impact factor of the journals they write in.

One thing our results show clearly, however, is that OSS projects are not homogenous datasets with a cohesive identity or strongly defined set of behaviours. Indeed, others have noted that projects have individual release cycles, development norms, and needs[9,41]. Casari, Ferraioli, and Lovato point out that OSS is a *sociotechnical* system, in which researchers often forget to address the "*socio-*" element of the system, since the technical aspects are much easier to access and assess[83].

In "Ten simple rules for funding scientific open source software", an article co-authored by research funders from the Chan Zuckerberg Initiative and the Sloan Foundation, Vu et al. assert that projects of different ages and maturity levels will have different needs, and can not be assessed on a simple set of metrics. The findings in this research study fully support this assertion: **health and activity indicators will have different meanings over time, depending on project-specific practices, the age, and status of the project**. For example:

**Number of paid staff in the past:** If this number went up significantly, it could have indicated multiple types of crisis - perhaps, running out of funding or serious project culture issues driving people away. At the same time, older projects by their nature will have staff turnover, and this might not be a sign for worry at all.

**Downloads or commits over time** can be a misleading indicator. Granular commits coupled with complex automated workflows can inflate commit statistics to appear "busier" than other repositories.

For example, if a modern development project uses continuous integration (CI) tooling, every developer commit might result in a re-download of a package used as a dependency, when the CI re-builds the project with the newly added computer code. For an active developer, this could possibly even happen multiple times an hour. A relatively small or young project that is bundled as a dependency could easily get a high download count because of actions of its own developers, and appear more "successful" than a similar project that does not function as a downloadable dependency. Project 79 was one concrete example of a project in this study with a highly regimented commit workflow that resulted in thousands more commits compared to other projects.

## Conclusion

No projects failed during the course of the study, meaning we are unable to draw strong conclusions about what project attributes are *essential* to sustainability. Very little was common amongst the projects, beyond an open ethos, expectations that they would (or should) continue for at least a year after the starting point of the study (intent to continue into the future), and a project leader who felt they had sufficient authority to participate in the study. With further exploration, it might be possible to ascertain if empowered leadership or longer-term intent are essential attributes for sustainability, but that is beyond the scope of this study.

The highly heterogeneous nature of the projects, however, provides a wealth of evidence about the attributes that *are not essential to project longevity*, and should not be treated as predictive. Presence of licence files, contribution guidelines and code of conduct all varied across our sample, as did mean time to resolve pull requests and issues, number of paid staff and leadership, and direct grant funding.

We conclude by noting that whilst the appeal of paste-repository-url "open source scoreboards" is high, the risks are probably higher. Automated repository-based tools do not - and may never have - the sophistication needed to understand all the different caveats of a software project lifecycle, and are attempting to draw conclusions from only part of the data - the part which is most easily observed.

It would be unfair to any open source maintainer to create a system where the search for a quick-and-easy assessment method oversimplifies their work and completely devalues or ignores the complex interpersonal elements needed to build thriving communities. In order to effectively understand longevity and sustainability of OSS, it is imperative to create a <u>qualitative and contextualised understanding</u> of the project's goals, plans, governance, development behaviour, and culture, as **version control metrics and indicators alone will never tell a full story**.

## Future work

The process of running this study provided ample material for future work. In particular, it would be helpful to see a similar study over a longer period of time - three to five years, or even longer. This would allow time for questions such as "where do you see your project in five years" to be compared to empirical outcomes.

More granular funding- and project-structure questions might provide additional context to research software sustainability, as well.
What types of funding support does a project have at the start and throughout the study? Perhaps it is industry-based support, grant funding from agencies, in-kind staff time from PIs, contributors, or students? Academia is notorious for its staff precarity[82] - so how precarious are the jobs of project PIs and other contributors?

"Hidden" work in open source projects and academic projects is a known challenge[82]. How many people are involved in open source research projects, but don't leave easily detectable traces in their source code repositories? At least one of our respondent project leaders had never committed a single line to the repository link that they provided. Do teams with these non-code roles have more success in terms of longevity, adoption, or funding acquisition? Does the PI have an academic skill-set focused on papers and citations, and do they have people-management skills or coding skills as well?

Roles like a team administrator, a user experience designer, a good people manager, or a community manager might not have much direct effect on the codebase, but might be significant when it comes to project success. Even if these aren't official roles, does anyone in the team have these skills, or act in this capacity under another job title?

In addition to looking at the funding and staffing makeup of academic / research software teams, it would be interesting to look at other communities with similar behaviours. This might include commercial and industry-sponsored open source software, inner-source software (software that is NOT open source, but uses open source collaboration paradigms in their development behaviour[84]), and other open communities that use version control social platforms (such as GitHub) for collaboration[85]

# Appendix

## Methods: Additional information

### OpenRefine anonymisation example

We manually anonymised the data transform history metadata after the primary data had been anonymised. An illustrative example of the steps is shown in the table below - changes are highlighted in yellow.

| | | EXAMPLE |
|---|---|---|
| 1:<br>Data cleaning | Data tidying and normalisation | Github.com/Manchester/SomeRepo<br><br>becomes<br><br>https://github.com/manchester/somerepo |
| 2:<br>Anonymisation | Primary data | "University of Manchester"<br><br>becomes<br><br>"[Research Institute]" |
| 3:<br>Anonymisation | Data transform history metadata | {from: "I received funding from Manchester", to: "I received funding from [Research Institute]"}<br><br>becomes<br><br>{from: "STEP REMOVED DUE TO ANONYMISATION REQUIREMENTS", to: "I received funding from [Research Institute]"} |

### Timeline

Response periods for each survey were open for around six weeks, as shown in the timeline table below:

| Year | 2021 | 2022 |
|---|---|---|

| Month | M | J | J | A | S | O | N | D | J | F | M | A | M | J | J |
|---|---|---|---|---|---|---|---|---|---|---|---|---|---|---|---|
| Month 0 start point survey *(2021-05-03 to 2021-06-21)* | ■ | ■ | | | | | | | | | | | | | |
| Month 6 mid point follow up *(2021-11-18 to 2021-12-21)* | | | | | | | ■ | ■ | | | | | | | |
| Month 12 end point follow up *(2022-05-31 to 2022-07-14)* | | | | | | | | | | | | | | ■ | ■ |

**Table 11**: Timeline for administering the survey component of the study.

## Results: Data Annexe

This data are a summary of all indicators which had complete results-sets, and includes a presentation guide to aid interpretation of graphs, as well as highlighting individual interesting data points and case studies within each section. Key data has been highlighted within the "results" section of this paper.

### Data presentation guide

Visualising data that changes over time can result in complex graphs. While we have made our best effort to present data as clearly as possible without oversimplifying, this section explains some of the data presentation methods with illustrative examples.

### Single-answer questions

Throughout this section, we present data from all three time points side by side. **Tables** summarising respondent responses to **radio-button single-answer questions** usually consist of an answer column, and three time-point columns, as illustrated below:

| **Answer** | **M0** (Month 0) | **M06** (Month 6) | **M12** (Month 12) |
|---|---|---|---|
| Total # of respondents | 38 | 34 | 34 |
| at this time point | | | |
| Response 1 | # of respondents who selected **Response 1 in month 0 (**% of respondents shown in brackets) | # of respondents who selected **Response 1 in month 6** (% of respondents shown in brackets) | # of respondents who selected **Response 1 in month 12** (% of respondents shown in brackets) |
| Response 2 | # of respondents who selected **Response 2 in month 0** (% of respondents shown in brackets) | # of respondents who selected **Response 2 in month 6** (% of respondents shown in brackets) | # of respondents who selected **Response 2 in month 12** (% of respondents shown in brackets) |

**Table 12**: Example data table to illustrate data presentation format.

Survey: Multiple choice questions

Some tables contain answers to check-box **multiple choice questions**. In this case, respondents may choose more than one option. The total number of <u>respondents</u> is still shown at the top of the table, but note that each individual respondent may have chosen more than one option, so summing up the count of responses to each question may add up to a number higher than expected. Illustrative example below using mock data - in month 0, summing response 1 and response 2 adds up to a count of 48 responses, even though we only have 38 total respondents. This is because 10 respondents chose BOTH response 1 AND response 2.

| **Answer** | **M0** (Month 0) | **M06** (Month 6) | **M12** (Month 12) |
|---|---|---|---|
| Total # of respondents at this time point | 38 | 34 | 34 |
| Response 1 | 20 | 18 | 21 |
| Response 2 | 28 | 17 | 16 |

**Table 13**: Example data table for multi-choice questions, where the total of responses may add up to more than the total number of responses

Visualisations

In some cases, we have presented graphs which visualise responses from each individual project over time. These graphs are presented as bar charts, with a structure similar to the table format above. Project names are shown on the left, followed by three colour-coded bars representing a project's response at each of the three time points. These graphs are sorted either by project founding date (older projects at the top, younger projects at the bottom), or by a project's self-reported size at month 0. Illustrative example, using real data, shown in the figure below:

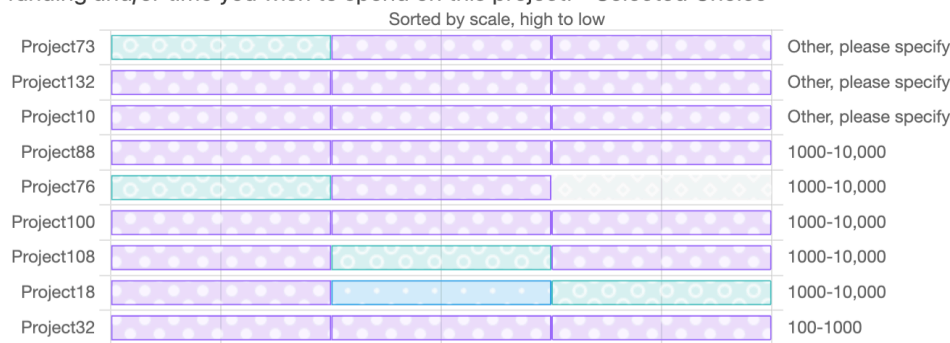

(... graph middle section truncated to save space)

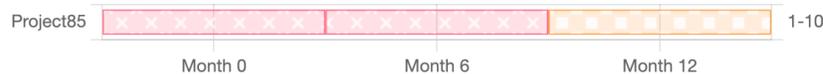

**Figure 5**: Example of three-survey data visualisation, showing responses at an individual project level over time.

In the figure above, moving left-to-right, the first column with coloured bars represents responses from each project in month 0, the middle column represents responses in month 6, and the right-most coloured bar column is month 12 responses.

To illustrate the flow of time descriptively, Project 85, one of the smallest projects shown in this graph, predicted in months 0 and 6 that it would be completely inactive in one year, but in month 12 (one year after month 0), they reported that they thought the project still might get occasional updates in the future. Projects 132, 10, 88, 100, and 32 all thought their project would still be active in one year's time, and did not see themselves leaving the project. This stayed consistent throughout all three timepoints.

## Side-by-side tables and graphs

In some cases, two graphs or tables were directly comparable. There were two primary circumstances where we did this:

1. Survey questions that were close variations of one another. e.g.
   a. "Where do you see your project in **one** year?" and
   b. "Where do you see your project in **five** years?",
2. The same data, sorted differently (usually by age or by project size as reported by the respondent at month 0).

In these cases, we presented these comparable tables and figures side-by-side, with the name of the variables / questions asked at the top of the graphs. Other than this, these visualisations and tables should be interpreted the same way as shown above.

## Open to contribution

**Survey question:**
Is this project (*still\**) open to contributions from the public?

> *the word "still" was included in months 6 and 12 only*
>
> **Source control assessment:**
>
> Is there an OSI licence, a CC licence, or some other open-style licence?
>
> Are there instructions for how interested parties can contribute?

Open to contribution: Survey

| Answer | M0 | M06 | M12 |
|---|---|---|---|
| Total # of responses | 38 | 34 | 34 |
| Yes | 37 (97.37%) | 34 (100%) | 33 (97.06%) |
| No | 1 (2.63%) | 0 | 1 (2.94%) |

**Table 14**: Number of open projects at each stage of the survey

37 of the 38 projects reported that they were open to contribution from the public in month 0, all responding projects were open to the public in month 6 (including the project that reported it was not open in month 0), and all but one were open to the public in the final survey.

Open to contribution: Source control: Licences

*Spelling note: Internationally, when using license/licence as a noun (not a verb), some countries use "license" and others use "licence". In this text, we use the form "licence", unless copying from another source, such as a licence's proper name, a result returned by an API, or a survey response.*

For the purposes of this study, we use the presence of a licence as a weak proxy for intent to allow others to collaborate on an open project. An open source (OSI-Approved) or Creative Commons licence generally allows others to download, re-use, and modify the work, which does not automatically mean the project is open to contribution from the public. Nevertheless, a total lack of licence generally means that under international copyright law, others can *not* download, re-use, or modify the work[22,86]. This lack of licence, therefore, <u>may</u> indicate a creative work or computer code author does not intend or wish to collaborate openly.

All participating repositories, bar one, had OSI-approved, Creative Commons, or custom open licences.

| Licence type | # of projects | Licence type (ctd) | # of projects |
|---:|---|---:|---|
| MIT | 9 | LGPL-3.0 | 1 |
| Apache-2.0 | 8 | Custom, Permissive | 1 |
| BSD-3-Clause | 7 | CC0-1.0 | 1 |
| * GPL-3.0 | 7 | MPL-2.0 | 1 |
| NOASSERTION | 6 | * AGPL-3.0 | 1 |
| * GPL-2.0 | 4 | None | 1 |
| * CC-BY-SA-4.0 | 4 | Artistic-2.0 | 1 |
| CC-BY-4.0 | 3 | BSD 3-Clause License | 1 |
| BSD-2-Clause | 2 | BSD-3-Clause-Modification | 1 |
| Perl 5 | 1 | | |

Table 15: Licences of participating repositories. Copyleft style licences are indicated by an asterisk.

In table 20 above, "NOASSERTION" is a result returned by GitHub for repositories that had no licence GitHub could find. Originally, sixteen of the results returned reported that they had no licence. After manual review, we were able to add ten more licences to the list. These were either licences that were not in the standard location for a licence (e.g. perhaps they were on the project website, or buried in a sub-folder in the repository), or they were non-standard licences with an open intent behind them.

One project - Project85 - had no licence at all. The other "NOASSERTION" repositories were all from respondents that had provided multiple source control URLs, at least one of which *DID* have a licence.

## Open to contribution: Source control: Contributing guidelines

Perhaps a stronger signal (than licence) for intent to collaborate openly with others is when a repository provides clear instructions for how newcomers can contribute to the repository.  In Month 0, 18 projects in total had CONTRIBUTING files in the root of their version control repository, as reported by GitHub's community API endpoint. We reviewed the "false" results manually and discovered that five of the repositories had contributing guidelines that were easily discoverable (using GitHub's embedded UI search function, searching for the terms "contribute" or "contributing") in other locations. Four were located in a sub-folder dedicated to project documentation, and one additional set of contributing guidelines were located within the project code, but not in the documentation folder.

This attribute did not appear to change significantly over the period of the study - only one project added contributing guidelines, and no projects removed them.

| Contributing guidance found? | M0 | M06 | M12 |
|---|---|---|---|
| True (Contributing file in root) | 18 | 18 | 19 |
| False | 14 | 14 | 13 |
| Docs | 4 | 4 | 4 |
| Search | 1 | 1 | 1 |

Table 16: Number of projects with guidelines dedicated to welcoming new contributors.

### Individual points of interest

One project (Project 33) reported that it began to accept contributions during the course of the study, one project (Project 85) ceased to accept contributions, and all the other respondents were open to public contributions throughout the whole study period. Both were less than one year old at the time of the initial survey.

*Project 85 - a short-term project with a planned sunset*

Project 85 reported at the start of month 0 that they expected the project to be wrapped up in one year's time, so month 12's results (that the project was no longer accepting contributions) are consistent with their initial expectations. Project85 was also the sole repository that did not have a licence file.

*Project 33 - mirrored open repository, with an access-restricted "core"*

Project 33 provided context for why they were not initially accepting contributions:

> "We maintain two GitHub repositories for our project, one through an [Anonymised] Enterprise account, and another through a public account (link shared as part of survey responses). We had originally started with an [Anonymised] Enterprise account without realizing that we would never be able to make our repository public, so we created a separate repository through a public account to ensure our materials would be open. When we make changes to our files, we push to both the Enterprise and public accounts, so the two repositories are essentially mirrored. However, we use the Enterprise account for project management - i.e., we post issues, have dev branches, make pull requests, etc., on the Enterprise account, and none of this is seen on the public account. I believe the only way to get access to our Enterprise account would be to have an [Anonymised] email address."

Project 33's non-enterprise GitHub repository shows granular commits from five Github users, one issue, and does not show any pull requests. Commit messages suggest that GitHub issues and pull requests are used as described, in a private enterprise account that is not visible in the public github account. Example commit messages: "`Merge pull request #22 from [ANONYMISED private repository name]/develop`" and "`fixed #97`". (The exact issue / pull request numbers have been changed to maintain anonymity).

Project 33 is a software project, but does not have an appropriate software licence. Instead, it has a Creative Commons licence, despite the fact that Creative Commons specifically asserts that their licences are not suitable for computer code[87].

## Project leadership: Respondent pay and permanence

**Survey question:**

Is this project part of your job? Are you paid to work on this project? Select all that apply.

9 participating project leaders were entirely unpaid to work on their project at each time point in the study. This outnumbered individuals who had permanent contracts at all time points - less than one quarter of the respondents had permanent contracts to work on their project during the course of this study.

| Answer | M0 | M06 | M12 |
|---|---|---|---|
| Respondents | 38 | 34 | 34 |
| No answer | 0 | 0 | 0 |
| No | 9 (23.68%) | 9 (26.47%) | 9 (26.47%) |
| Other, please specify | 6 (15.79%) | 7 (20.59%) | 9 (26.47%) |
| Yes, as a staff member with a permanent contract | 7 (18.42%) | 6 (17.65%) | 8 (23.53%) |
| Yes, as a student | 2 (5.26%) | 1 (2.94%) | 1 (2.94%) |
| Yes, as a staff member with a temporary or fixed-term contract | 14 (36.84%) | 12 (35.29%) | 11 (32.35%) |

Table 17: project leader pay / employment arrangements

Two students responded to the study - one maintaining their student status throughout the whole study (Project 34), while the other (Project 10) presumably graduated between month 0 and month 6. Both student project leaders provided additional context around their answers in the "anything else" free text field at the end of the survey.

Individual points of interest

| | | M0 | M06 | M12 |
|---|---|---|---|---|
| Project 10 | you-paid | Yes, as a student | No | Yes, as a staff member with a temporary or fixed-term contract |
| Project 10 | anything-else | Funded by two [Government funder] doctoral studentships | For the user id I've provided mine and the organisation that owns the repo ([Anonymised]). | As this is academic software it is mostly maintained by PhD students. In our case we work collaboratively with an industry partner. In the future you may want to ask about third-party sponsorship. |
| Project 34 | you-paid | Yes, as a student | Yes, as a student | Yes, as a student |
| Project 34 | any thing-else | For myself and the other people I listed as paid to work on the project -- we're all grad students (and one postdoc), none of us hired to explicitly work on Project34. We're working on it because it's related to our research, so we figured it was worth the effort to go beyond just writing research code and make it more general. But it's in no one's job description and **will be part of no one's dissertation.** | Determining if I'm paid / if others are paid for the project is hard. Currently, 6 of the 7 contributors are PhD students. **I'm the only one currently planning on putting it in my dissertation** (and hence I counted myself as being paid to work on it), but that might change. I'm not planning on seeking funding, but I have an informal arrangement with my advisor for a post-defense job that will involve me working on this project. | |

Table 18: Answers from project leaders (who were also students) about their pay/employment status

Other respondents, such as Project 132, gave "other" answers that indicated they were students in the free text, but that their work as a student did not give them explicit permission to work on their project, so they didn't fully count themselves as paid:

| | | M0 | M06 | M12 |
|---|---|---|---|---|
| Project 132 | you-paid | Yes, as a staff member with a temporary or fixed-term contract | Sort of, my work [on] the project is tolerated at my day job | Currently not in the job description, will get a second job for this after summer, then will become part of my job description after I finish my PhD |

Table 19: Project 132 answers about their pay/employment status

People paid to work on the project: Paid staff

> **Survey questions:**
>
> Have others been paid to work on the project **in the past**? If so, how many? This does not include interns.
>
> How many others are **currently paid** to work on this project? This does not include interns.
>
> **Indicator:** Grant funding

| Survey responses - paid staff - sorted by project **age** | |
|---|---|
| Paid staff in the **past** | **Current** paid staff |

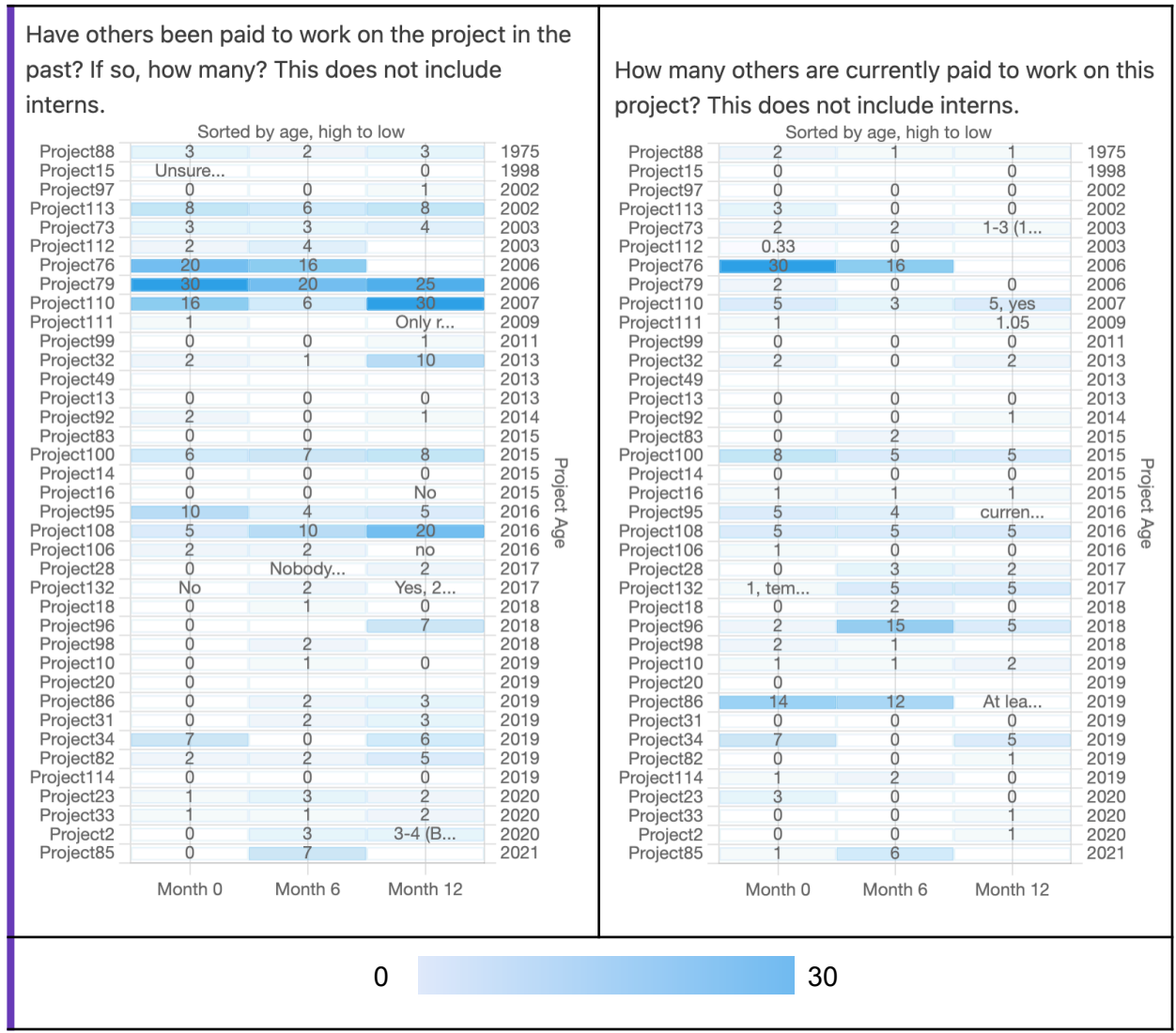

Figure 6: Paid staff per project in the past (left), and at the time of response (right), *sorted by age*

| Survey responses - paid staff - sorted by project **size** | |
| --- | --- |
| Paid staff in the **past** | **Current** paid staff |

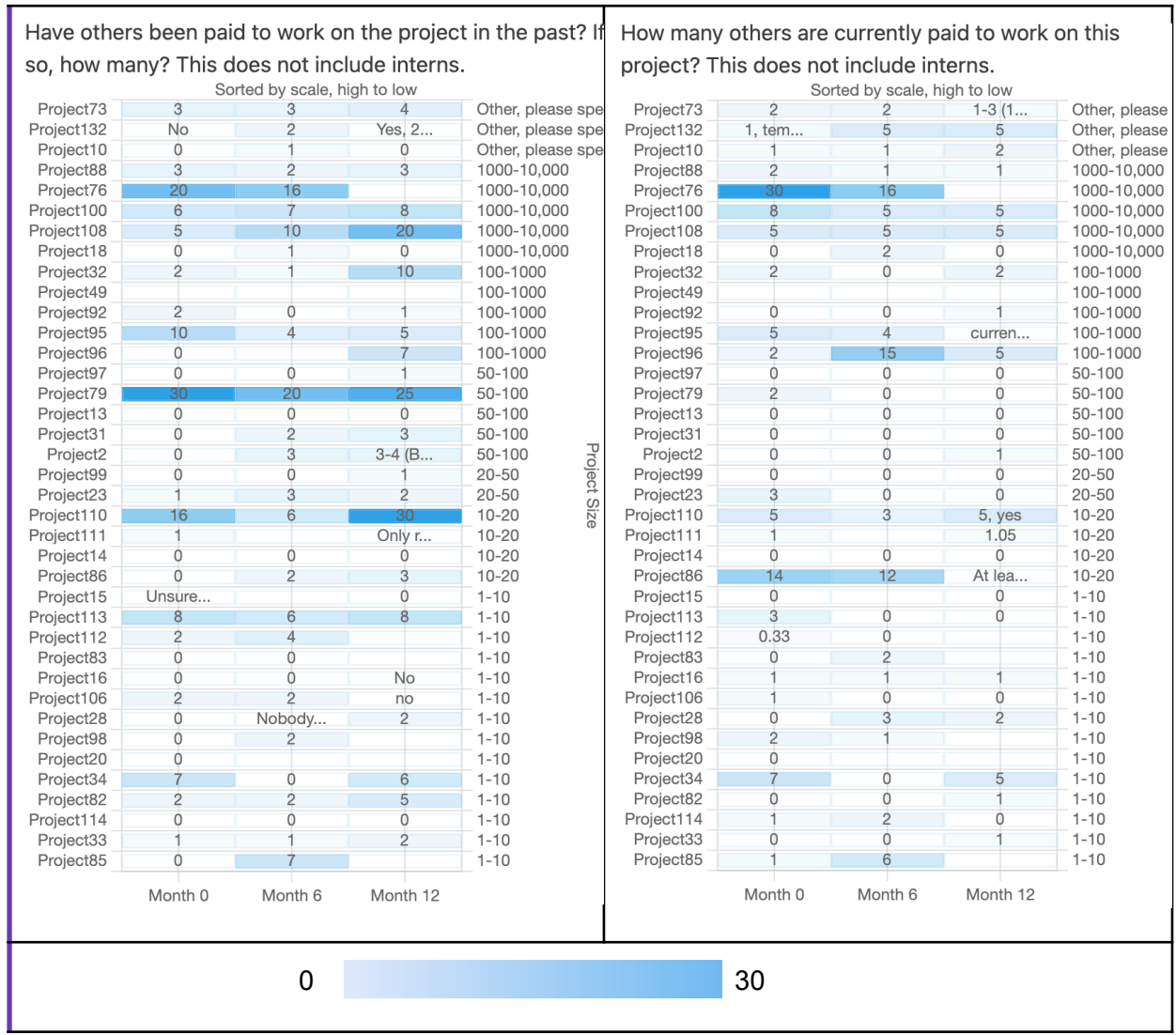

**Table 20**: Paid staff per project in the past (left), and at the time of response (right), sorted by *project size*

Projects younger than five years old at the start of the survey (i.e. founded in 2016 or later) never reported more than ten former staff members.

### Individual points of interest

Being an older or larger project did not automatically mean that projects had paid staff. Some of the very oldest projects (such as project 15 and project 97, both more than 20 years old at the time of writing) did not report having paid staff for the majority of the study period. Similarly,

Project73 - the largest project that participated in this study - had far fewer current or former staff than many of the other smaller projects.

| | Have others been paid to work on the project **in the past**? If so, how many? This does not include interns. | | | How many others are **currently** paid to work on this project? This does not include interns. | | |
|---|---|---|---|---|---|---|
| | **# of paid staff** | | | **# of paid staff** | | |
| **Project** | **M0** | **M06** | **M12** | **M0** | **M06** | **M12** |
| **Project 15** | Unsure about this, though the tools in this project had been used extensively (in one form or another) during the early days of genome sequencing and analysis, including the Human Genome Project, so there is very likely code that benefitted from grant funding. | 0 | 0 | 0 | 0 | 0 |
| **Project 97** | 0 | 0 | 1 | 0 | 0 | 0 |
| **Project 73** (estimated >1 million users) | 3 | 3 | 4 | 2 | 2 | 1-3 (1 full time, several people are able to spend work hours on it) |

Table 21: projects with low numbers of staff at different maturity levels

People paid to work on the project: Interns

**Survey questions:**

Have you ever had paid interns working on this project, e.g. through Google Summer of Code, Outreachy, or some other internship scheme? How many? *Please enter 0 if you have had no interns, or make your best guess if you are not sure. If any interns are working right now, please include them in the total.*

**Indicator:** Mentorship

---

**Survey responses - interns**

| Interns - sorted by project **size** | Interns - sorted by project **age** |
|---|---|

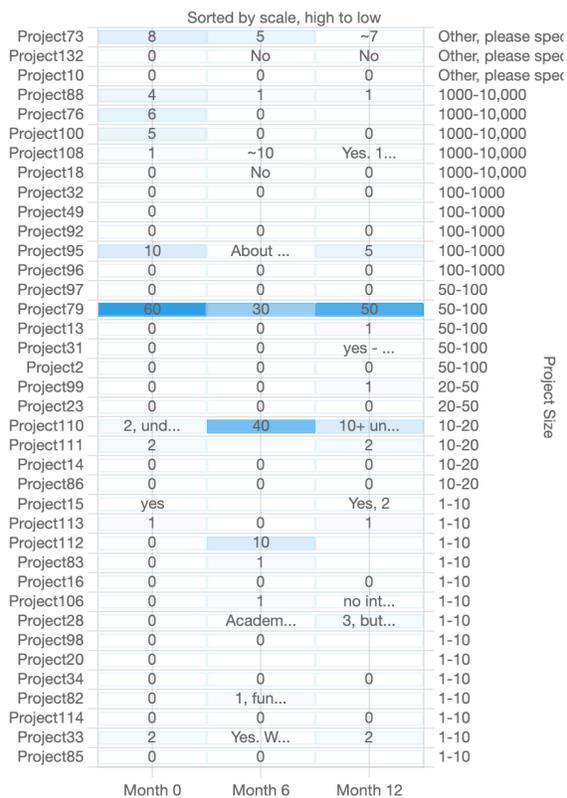

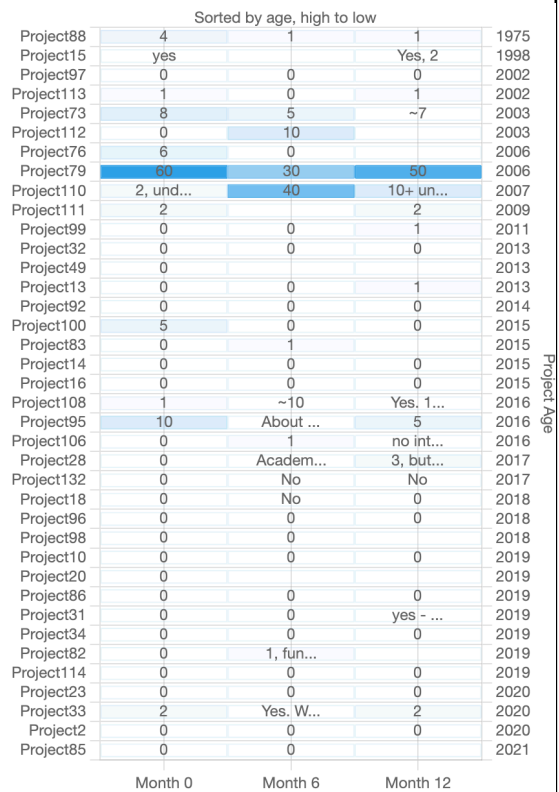

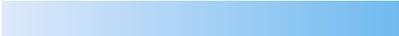

| 0 | | 60 |
|---|---|---|

Figure 7: Number of current or past project interns, sorted by project size (left) or project age (right).

Most projects did not have support from interns, and the results did not appear to cluster around age or project size. Most projects did not show evidence of sustained intentional mentorship

Project Culture: Behavioural expectations in codes of conduct

**Source control question:**

Is there a Code of Conduct file present in the source control repository, or other behaviour guidelines?

**Indicator:** Code of conduct

Some projects use Codes of Conduct to set behavioural standards for their community[37,49,53,88]. Similar to the contributing and licence detection, GitHub's API attempts to automatically detect whether or not a project has a code of conduct, but does not always report correctly if these files are outside of expected locations. We manually checked for codes of conduct in repositories that reported there was no code of conduct, and found two more by typing "conduct" into the GitHub UI search engine - one in the .github folder, and one slightly more deeply nested in the code. Both older (and therefore longer-sustained) projects as well as younger projects had similar spreads of codes of conduct - around two thirds had the file present in some form, and the remaining repositories did not.

| Code of Conduct found? | Month0 | Month12 |
|---|---|---|
| false | 19 | 17 |
| true | 16 | 18 |
| search | 1 | 1 |

| .github | 1 | 1 |
|---------|---|---|

**Table 22**: Codes Of Conduct found in participating repositories

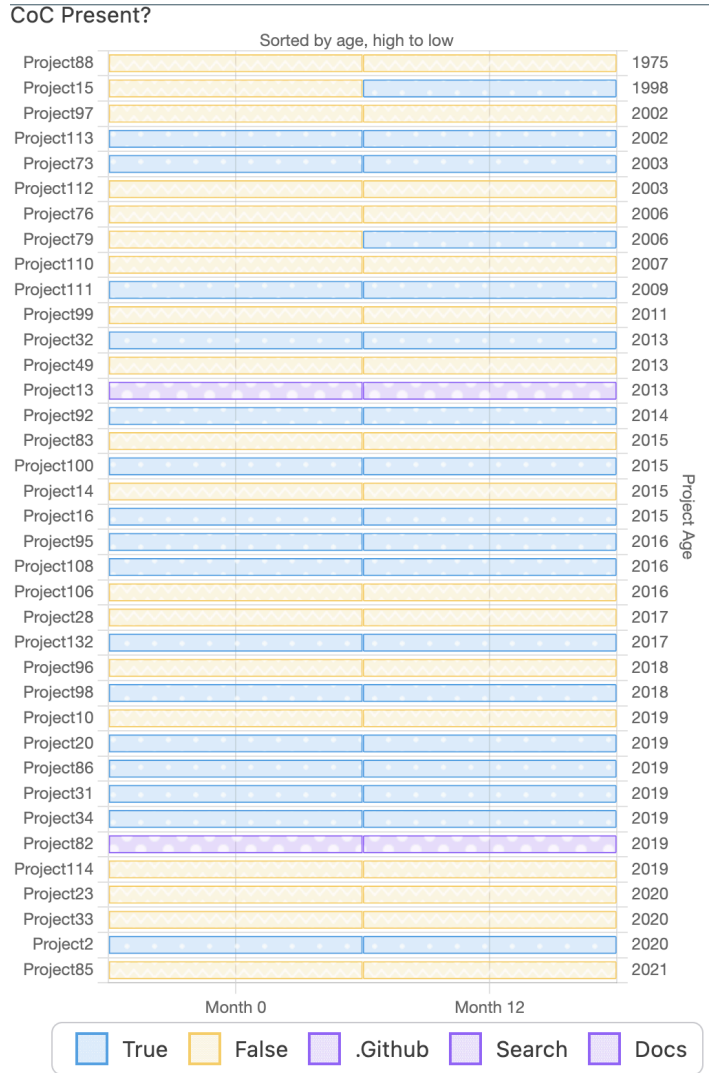

**Figure 8**: Presence of a code of conduct in a project's GitHub repository. As most projects did not change over the entire time period, we omitted Month 6 from the visualisation.

Continuity: hopes in an ideal-case scenario

> **Survey questions:**
>
> In an ideal situation, where do you see your project one year from now? Assume you have the funding and/or time you wish to spend on this project.

Given that some projects might be time-boxed or limited in some way, e.g. a planned end of a clearly scoped project, we did not assume all projects expect to run indefinitely[89]. We asked project leaders what they *would like to see in an ideal scenario,* if resources were not an issue.

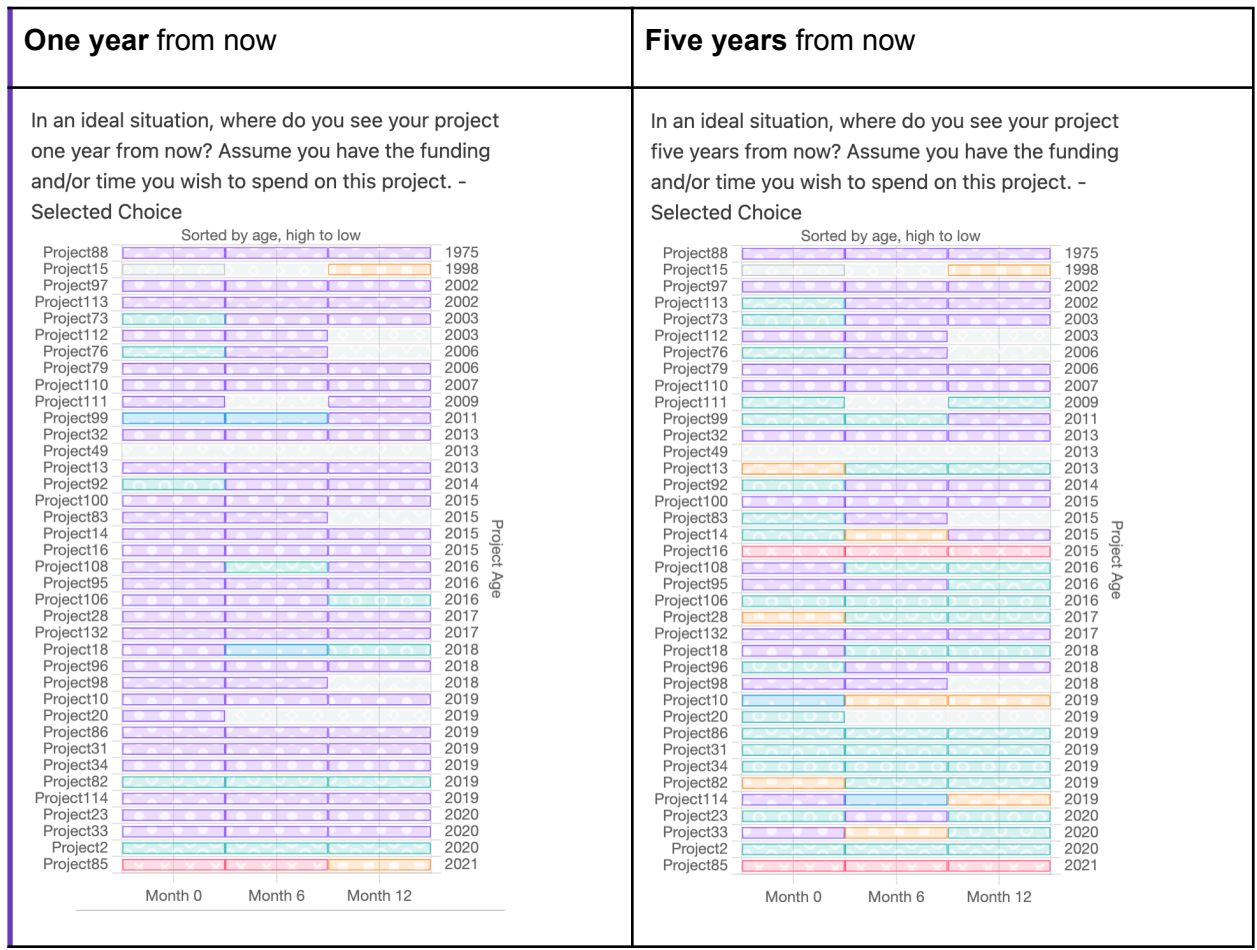

| **One year** from now | **Five years** from now |
|---|---|

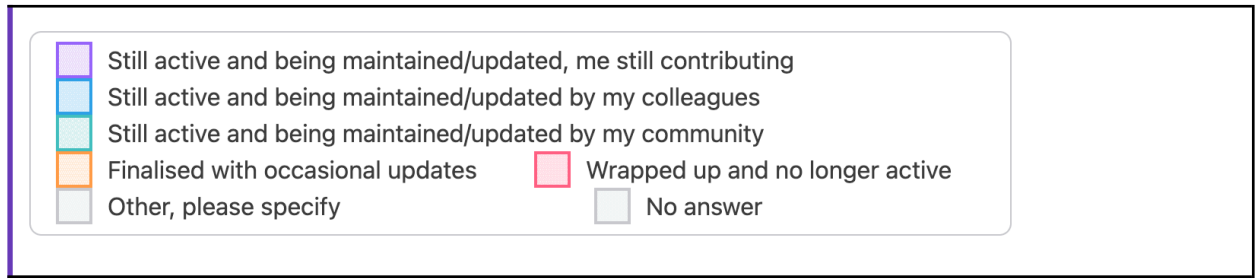

Figure 9: In an **ideal** situation, where do you see your project (one/five) years from now?

Throughout the study, most project leaders expected both the project to be around in one year, as well as themselves actively involved in the project. By contrast, in five years, most project leaders expected the project still to be actively maintained, but did not necessarily expect themselves to be a contributing part of the project.

Only two project leaders foresaw the project wrapping up or entering a "maintenance mode" within a year - Projects 85 and 15. When looking five years ahead, this rose to five projects. Answers to the five-year outlook were also less consistent between the same project in subsequent surveys, compared to the one-year outlook.

| | In an ideal situation, where do you see your project **one year** from now? Assume you have the funding and/or time you wish to spend on this project. | | | In an ideal situation, where do you see your project **five years** from now? Assume you have the funding and/or time you wish to spend on this project. | | |
|---|---|---|---|---|---|---|
| | **# of projects** | | | **# of projects** | | |
| **Answer** | **M0** | **M06** | **M12** | **M0** | **M06** | **M12** |
| *# of survey responses* | *38* | *34* | *34* | *38* | *34* | *34* |
| still active and being maintained/ updated, me still contributing | 29 (76.32%) | 28 (82.35%) | 27 (79.41%) | 14 (36.84%) | 17 (50%) | 14 (41.18%) |
| still active and being maintained/ updated by my colleagues | 1 (2.63%) | 2 (5.88%) | 0 | 1 (2.63%) | 1 (2.94%) | 0 |
| still active and being maintained/ updated by my community | 5 (13.16%) | 3 (8.82%) | 5 (14.71%) | 16 (42.11%) | 11 (32.35%) | 15 (44.12%) |
| finalised with occasional updates | 0 | 0 | 2 (5.88%) | 3 (7.89%) | 3 (8.82%) | 3 (8.82%) |
| wrapped up and no longer active | 1 (2.63%) | 1 (2.94%) | 0 | 2 (5.26%) | 2 (5.88%) | 2 (5.88%) |
| Other, please specify | 1 (2.63%) | 0 | 0 | 1 (2.63%) | 0 | 0 |
| No answer | 1 (2.63%) | 0 | 0 | 1 (2.63%) | 0 | 0 |

**Table 23**: One- and Five-year project status projections, as reported by project leaders.

Continuity: expectations if respondent is unable to maintain the project

**Survey question:**

If you find yourself unable to maintain the project any longer, what do you foresee happening? Select as many as apply to your situation

Most respondents were optimistic about their community's ability to continue with the work on the project, even if they themselves were not able to continue working on the project for some reason. Only one or two respondents at each survey point reported intent to close the project down if they themselves did not have time to maintain it. Over half of the respondents expected that others (community, colleagues, or staff) would continue maintaining the project, while 19 asserted they would still try to update the project, although possibly very rarely.

| Answer | M0 | M06 | M12 |
|---|---|---|---|
| Respondents | 38 | 34 | 34 |
| No answer | 0 | 0 | 0 |
| My community members would keep this project running. | 20 (52.63%) | 20 (58.82%) | 18 (52.94%) |
| My colleagues/employees would continue to work on this | 6 (15.79%) | 8 (23.53%) | 8 (23.53%) |
| I would continue to provide updates in my free time | 8 (21.05%) | 8 (23.53%) | 7 (20.59%) |
| I would provide periodic but rare updates when I could. | 11 (28.95%) | 8 (23.53%) | 11 (32.35%) |
| I would close the project down | 1 (2.63%) | 2 (5.88%) | 2 (5.88%) |
| Other, please specify | 1 (2.63%) | 1 (2.94%) | 0 |

Table 24: Project leader expectations for their project if they were unable to continue maintaining their project personally.

Respondents were able to select multiple choices, hence why the total number of responses is higher than the number of respondents.

Individual points of interest

The sole "Other" answer (Project 23) noted that while they believe their project is important and covers an otherwise unaddressed niche, they do not believe they would be able to maintain it if unfunded. Two illustrative quotes:

Project 23, Month0, in response to this question:

> We are a young project and I'm not aware of anyone who would be able to maintain it now, without funding, if I wasn't able. If I became unable now, I would offer it to anyone to take over and aim to provide input when I could. I hope we can move towards being maintainable by others.

Project 23, Month 6, in response to the final "anything else" free text prompt:
> We are committed to Project23, still believe it addresses an otherwise unmet opportunity, but are struggling to keep it funded.

## Survey questions - Month 0 - start point

| | Dataset variable | Question | Answer type | Answer options (if relevant) |
|---|---|---|---|---|
| 1 | screening-q | Are you leading or co-leading an open source / open science related project? This might mean you're a research software engineer who writes open source software, a leader of an open science community project or event, or a researcher who writes code in R, Python, or any other coding language, so long as the project is open. You don't have to be the PI of the project, but it's best if any given project only answers once. If you're unsure if this applies to you, please contact yochannah.yehudi@postgrad.manchester.ac.uk for clarification. | Boolean Yes/No | Yes, I lead an open source or open science project. No, I do not lead an open source or science project |
| 2 | consent-read-pis | I confirm that I have read the participant information sheet (on the previous page) for the above... | Boolean Yes/No | |
| 3 | consent-voluntary | I understand that my participation in the study is voluntary and that I am free to withdraw at any time without giving a reason and without detriment to myself. I understand that it will not be possible to remove my data from | Boolean Yes/No | |
| | | the project once it has been anonymised and forms part of the data set. I agree to take part on this basis. | | |
| 4 | consent -anon- publish | I agree that any data collected may be published in anonymous form in academic books, reports, journals, and data repositories. | Boolean Yes/No | |
| 5 | consent- regulator y | I understand that data collected during the study may be looked at by individuals from The University of Manchester or regulatory authorities, where it is relevant to my taking part in this research. I give permission for these individuals to have access to my data. | Boolean Yes/No | |
| 6 | consent- institutes | I agree that any personal/anonymised data collected may be shared with researchers/researchers at other institutions. | Boolean Yes/No | |
| 7 | consent - optional | Optional Consent item: You can still participate in the study even if you choose to answer "no" to the next item: I agree to share my pseudonymised survey responses openly online. (Optional). | Boolean Yes/No | |
| 8 | consent -overall | Overall agreement: I agree to take part in this study. | Boolean Yes/No | |
| 9 | consent -signed | Type your full name here to sign the form. | Text | |
| 10 | project-name | What is the name of your project? | Text | |
| 11 | project-founded | When was this project founded? It's okay to approximate, we don't need a precise day. | Text | |
| 12 | project-urls | What is your project's online code repository URL (e.g. GitHub URL). If you have more than one repository for your project please share as many URLs as you think are relevant, one per line. | Text | |
| 13 | project-open-contrib | Is this project open to contributions from the public? - Selected Choice | Yes/ No/ Other | |
| 14 | project-open-contrib_3 _TEXT | Is this project open to contributions from the public? - Other - Text | Free text for "other" option | |
| 15 | project-user-count | How many users / contributors / community members do you have at the moment? Estimate if you're not sure. Draw from analytics, GitHub, etc. as appropriate. - Selected Choice | Radio | 1-10  (1) 10-20  (2) 20-50  (3) 50-100  (4) 100-1000  (5) 1000-10,000  (6) |
| 16 | Project-user-count_7_ TEXT | How many users / contributors / community members do you have at the moment? Estimate if you're not sure. Draw from analytics, GitHub, etc. as appropriate. - Other, please specify - Text | Free text for "other" option | |
| 17 | Project-user-potentl | How many users / contributors / community members would you realistically like to have in the future? Consider the size of your target audience - for example, if you're creating a community or software for a small niche interest, try to be realistic about the maximum number of people possible. - Selected Choice | Radio | 1-10 (1)<br>10-20 (2)<br>20-50 (3)<br>50-100 (4)<br>100-1000 (5)<br>1000-10,000 (6) |
| 18 | Project-user-potentl_7_TEXT | How many users / contributors / community members would you realistically like to have in the future? Consider the size of your target audience - for example, if you're creating a community or software for a small niche interest, try to be realistic about the maximum number of people possible. - Other, please specify - Text | Free text for "other" option | |
| 19 | Project-interns | Have you ever had paid interns working on this project, e.g. through Google Summer of Code, Outreachy, or some other internship scheme? How many? Please enter 0 if you have had no interns, or make your best guess if you are not sure. If any interns are working right now, please include them in the total. | Text | |
| 20 | you-github- | What is your GitHub (or other code repo) user id? | Text | |

| | | | |
|---|---|---|---|
| | user<br><br>We need this in order to correlate your responses with your activity on GitHub. | | |
| 21 | you-email | What is your preferred contact email address?<br><br>We will not share this with others - we only ask so we can follow up in the future with our 6- and 12- month follow up surveys. | Text | |
| 22 | you-role | What is your role in this project? (e.g. founder, maintainer, contributor, mentor, intern) | Text | |
| 23 | leadership-team-size | How many people, including you, are in the leadership team for this project? | Text | |
| 24 | you-paid | Is this project part of your job? Are you paid to work on this project? - Selected Choice | Multiple options permitted | Yes, as a student  (1)<br>Yes, as a staff member with a permanent contract  (2)<br>Yes, as a staff member with a temporary or fixed-term contract  (3)<br>No  (4) |
| 25 | you-paid_5_TEXT | Is this project part of your job? Are you paid to work on this project? - Other, please specify - Text | Free text for "other" option | |
| 26 | funds-grant-funds | Is this project linked to grant funding with an end date? | Boolean Yes/No | |
| 27 | funds-others- | If you leave your role, would you expect others to pick this project | Boolean Yes/No | |
| | pick-up | up? | | |
| 28 | funds-oth ers-now | How many others are currently paid to work on this project? This does not include interns. | Text | |
| 29 | funds- others- in-past | Have others been paid to work on the project in the past? If so, how many? This does not include interns. | Text | |
| 30 | future- funding -plans | Do you plan to seek funding for your project in the future? | Boolean Yes/No | |
| 31 | future- one-year | In an ideal situation, where do you see your project one year from now? Assume you have the funding and/or time you wish to spend on this project. - Selected Choice | Radio | still active and being maintained/updated, me still contributing  (1) <br> still active and being maintained/updated by my community (2) <br> still active and being maintained/updated by my colleagues  (3) <br> finalised with occasional updates  (4) <br> wrapped up and no longer active  (5) |
| 32 | future- one-year _6_TEXT | In an ideal situation, where do you see your project one year from now? Assume you have the funding and/or time you wish to spend on this project. - Other, please specify - Text | Free text for "other" option | |
| 33 | future- five-years | In an ideal situation, where do you see your project five years from now? Assume you have the funding and/or time you wish to spend on this project. - Selected Choice | Radio | still active and being maintained/updated, me still contributing  (1) <br> still active and being maintained/updated by my community (2) <br> still active and being maintained/updated by my colleagues  (3) <br> finalised with occasional updates  (4) <br> wrapped up and no longer active  (5) |
| 34 | future-five-years_6_TEXT | In an ideal situation, where do you see your project five years from now? Assume you have the funding and/or time you wish to spend on this project. - Other, please specify - Text | Free text for "other" option | |
| 35 | future-cant-maintain | If you find yourself unable to maintain the project any longer, what do you foresee happening? Select as many as apply to your situation - Selected Choice | Multiple options permitted | My community members would keep this project running.  (1) My colleagues/employees would continue to work on this  (2) I would continue to provide updates in my free time  (3) I would close the project down  (4) I would provide periodic but rare updates when I could.  (5) |
| 36 | future-cant-maintain_6_TEXT | If you find yourself unable to maintain the project any longer, what do you foresee happening? Select as many as apply to your situation - Other, please specify - Text | Free text for "other" option | |
| 37 | anything-else | Final question: Anything you'd like to add that we didn't ask about? | Text | |

Responsiveness: pull requests, issues, and time taken to resolve them

> **Source control assessment:**
> What was the mean time to resolve (close and/or merge) an issue or pull request?

Limitation: Measuring issues and pull requests posed a challenge when retrieving information from busier repositories, as GitHub's API limits the number of calls made over a period of time. This resulted in some gaps in the data, despite repeated attempts over time to retrieve the missing data. Projects with at least one data point for these indicators are shown in the following graphs, and projects with no data across all three timepoints are omitted entirely. We thus recommend interpreting these results with more caution than other sections.

Time to resolve at month 0 represents historical data from before the study began. Note that language around pull requests and issues varies, but the two types of interaction are treated almost identically by GitHub's API. An open issue can only be closed (or resolved), but pull requests are either merged into the original codebase, or closed without merging the new code into the original codebase. We do not analyse the differences between these types of interactions, and Kalliamvakou et al noted that GitHub's API did not always return information about merged pull requests correctly as of 2014[69].

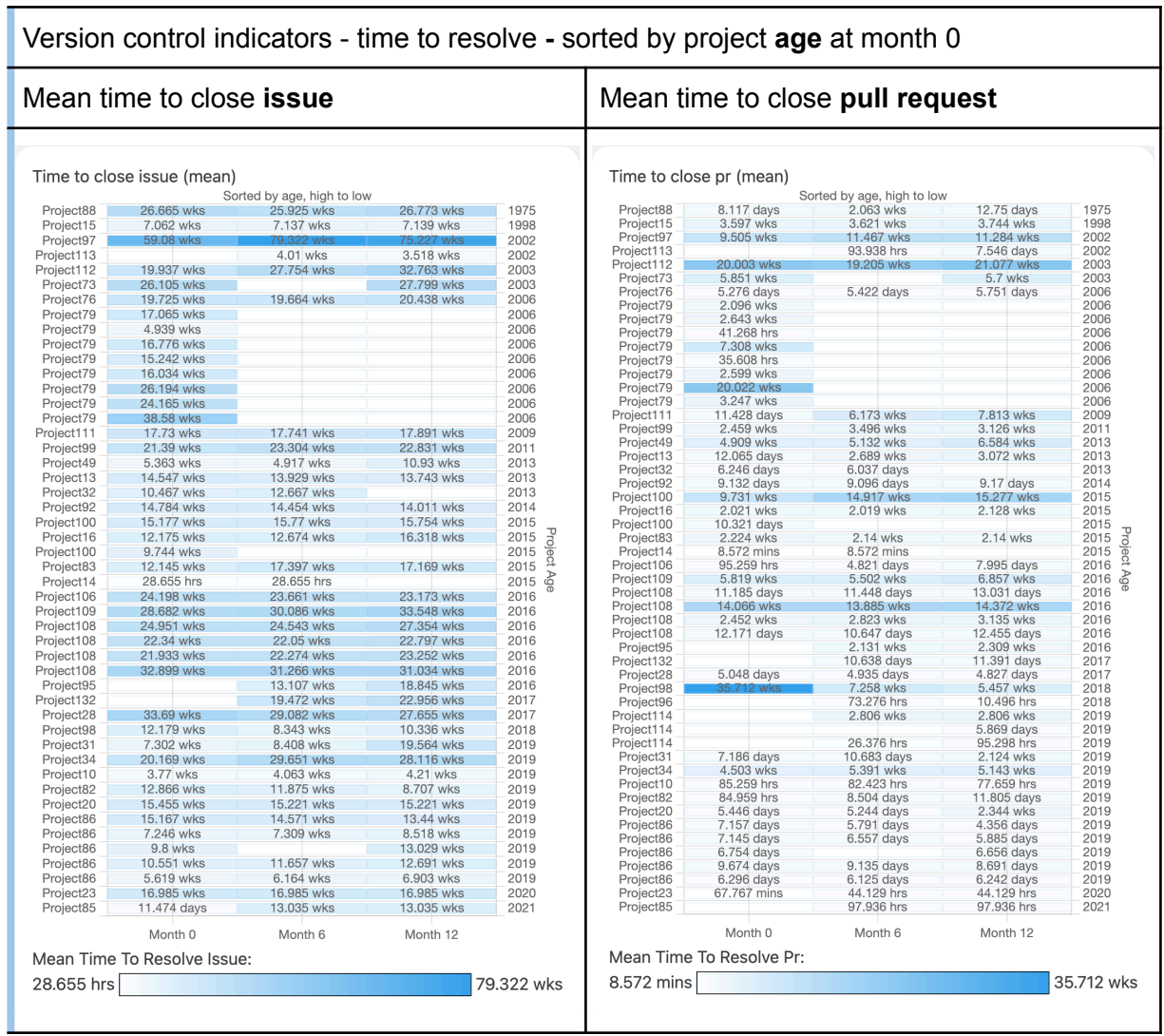

Version control indicators - time to resolve - sorted by project **age** at month 0

Figure 10: Mean time to close pull requests and issues on GitHub.

Time to close issues and pull requests were proposed in CHAOSS metrics as a way to measure project responsiveness, and to detect changes and biases in the community[90,91]. The metric specifications note that these measures are contextual and will differ depending on the types of action an issue or pull request might be addressing. Our results above seem to support this, as they vary both amongst projects and over time within the same project.

Individual points of interest

Older projects by default have the ability to have longer times-to-close, since they have, as they have been around longer. Nevertheless, some of the oldest projects, such as project 15 (founded in 1998), had relatively short times-to-close, with issues being consistently closed in around seven weeks, and pull requests in around three weeks. The issue close-time was similar to that of several of project 86's repositories - a project two decades younger, founded in 2019.

In this data set, the time taken to resolve a pull request is generally much faster than the time taken to resolve an issue, with the longest mean-time-to-resolve being 35 weeks for a pull request, but more than double this - 79 weeks - for issues. While we do not draw any conclusions from this point, we speculate in the discussion section on possible reasons for this pattern, and further avenues for investigation.

Size: number of commits

**Source control assessment:**

How many individual commits were made to a project's source control repository during the course of the study?

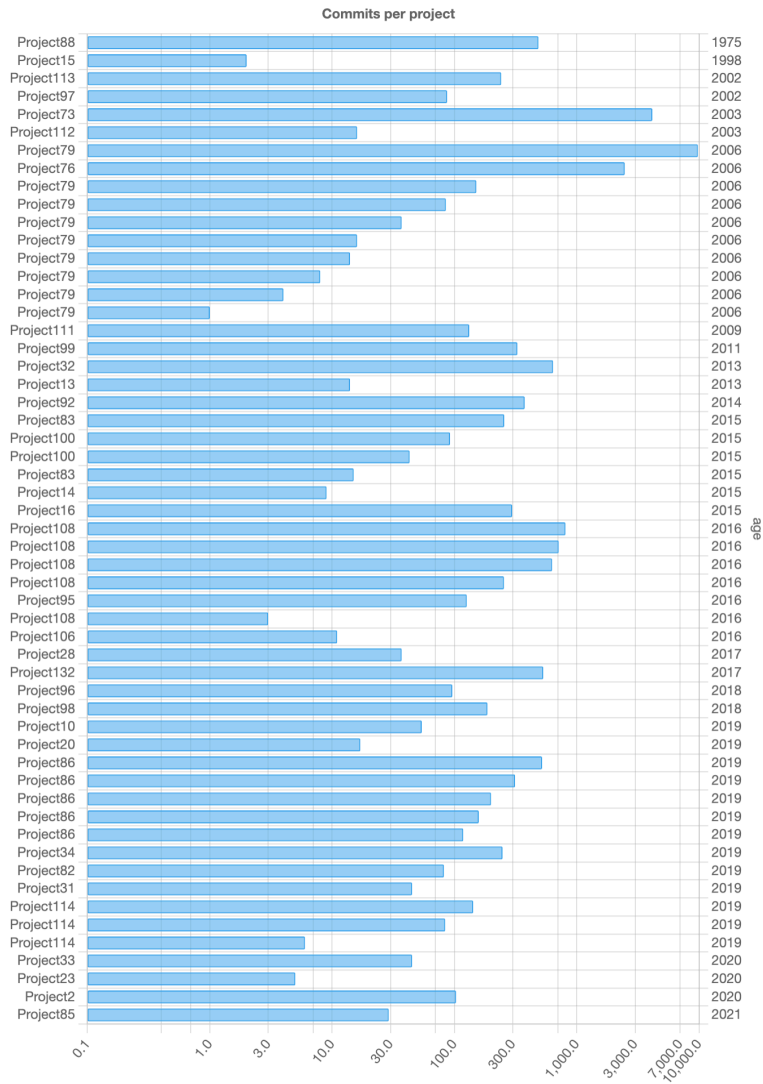

Figure 11: Number of commits made to individual repositories over the period of the study (log scale, sorted by project age). Projects that appear more than once submitted multiple version control repository urls.

Individual points of interest

Project 79, one of the older projects, not only had multiple repositories with a high number of commits, but also had over five thousand more commits than the next highest project, resulting in the need to show Figure 11 in log scale. Around 15% of the ~11000 commits in this repository were directly made by workflow management bots, and many of the other commits made by human users appeared to be of a similar package-management nature.

Size: current, and expectations/hopes for the future

**Survey questions:**

-   How many users / contributors / community members do you have *at the moment?* Estimate if you're not sure. Draw from analytics, GitHub, etc. as appropriate.

-   How many users / contributors / community members would you realistically *like to have in the future*? Consider the size of your target audience - for example, if you're creating a community or software for a small niche interest, try to be realistic about the maximum number of people possible.

Survey: project leader estimated project size

Project leaders were asked to estimate their project's current size, and their hopes for future growth. Small projects more often reported consistently across the three survey timepoints.

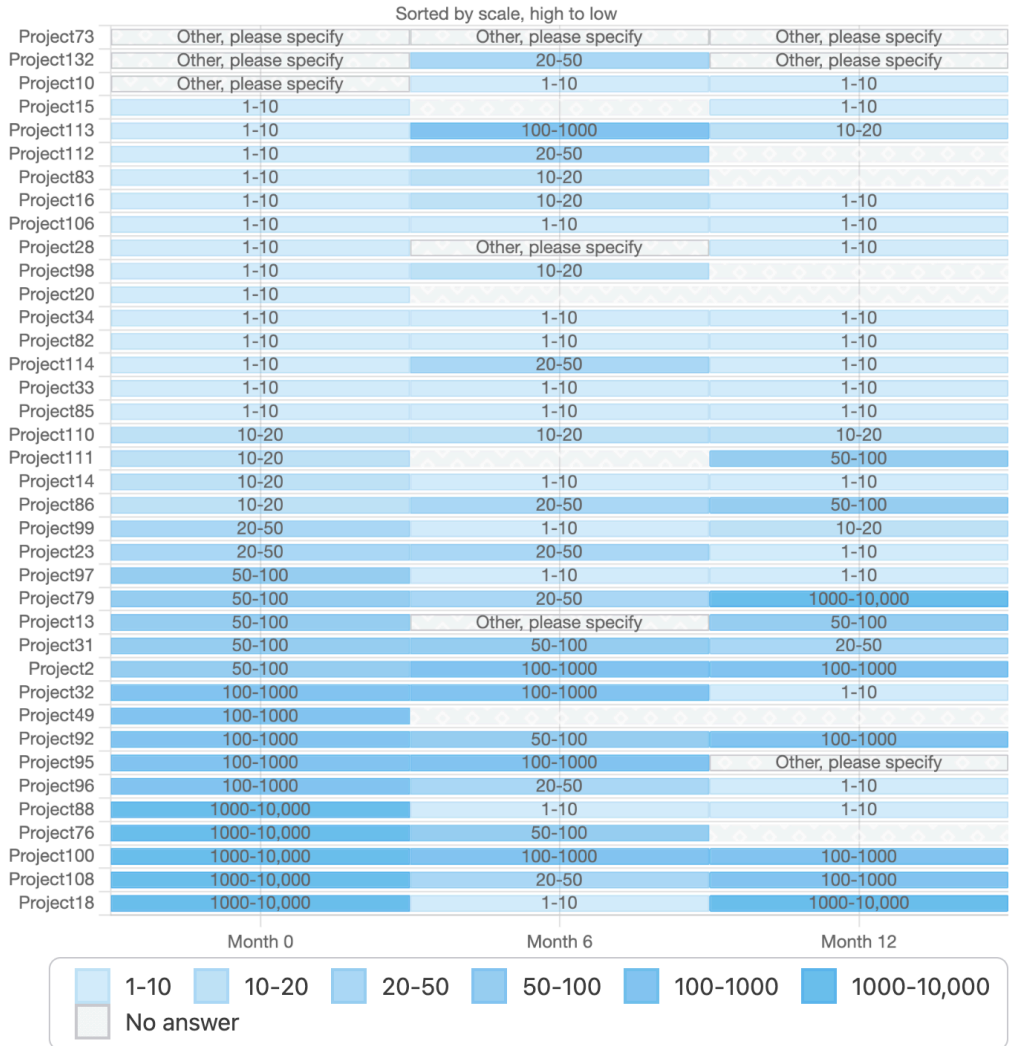

Figure 12: Projects' self-reported sizes throughout the study.

Given the fact that project structure varies[41], when we designed the questionnaire, we intentionally left the wording vague with regards to how to count project size - "users / contributors / community members". Several of the "other" responses were comments from respondents for whom the number of active contributors vs users was a very different number. See month 6 responses below from Project 73 for an illustrative response.

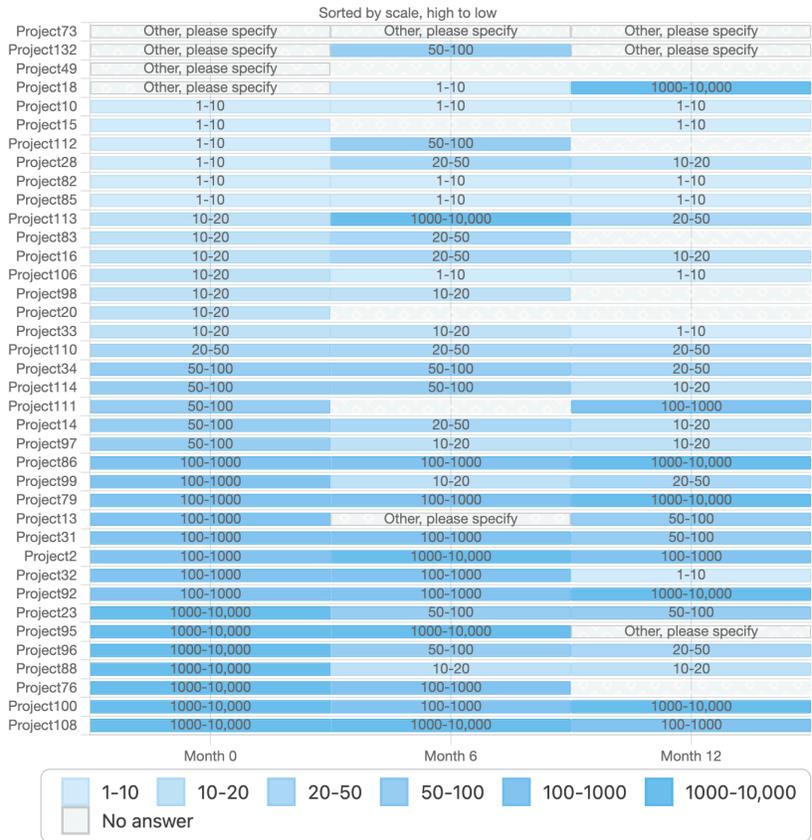

Figure 13: Projects' desired size throughout the study.

Individual points of interest

Project 73 was significantly larger than the survey scale; they reported around 150,000 users at month 0, growing up to over 1 million by the end of the study.

| Project 73 (Largest-scale respondent) | | | |
|---|---|---|---|
| | **M0** | **M06** | **M12** |
| Reported size | 150,000 | 1M+ users, few hundred contributors a year, 15ish regular contributors | 1M-5M |
| Desired size | 10,000,000 | 20 regular contributors | 10,000,000 |

**Table 25**: Responses from Project 23, the largest open source project that participated in this study.

Funding: During the study, and plans for the future

> **Survey questions:**
>
> Is this project linked to grant funding with an end date?
>
> Do you plan to seek funding for your project in the future?

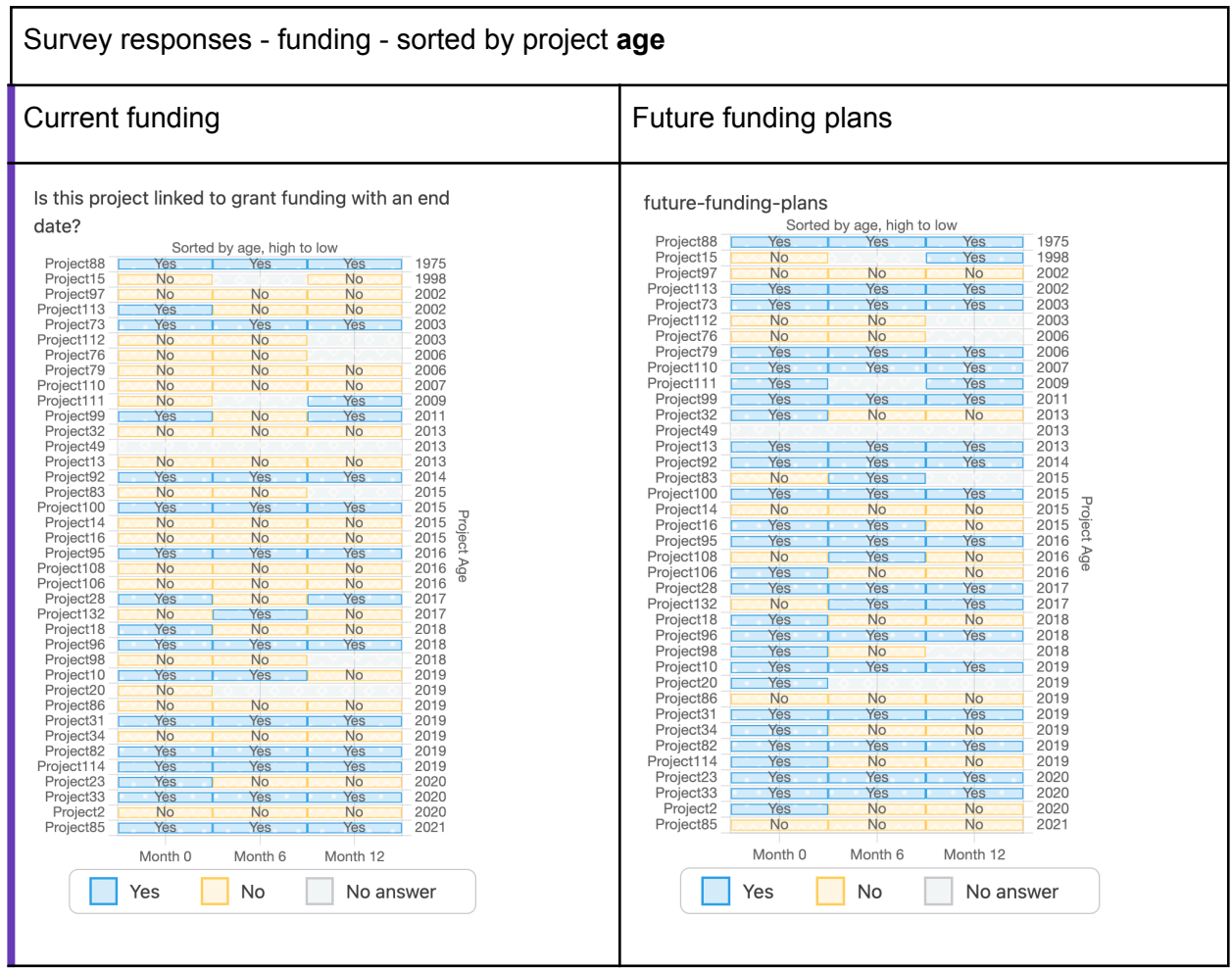

Figure 14: Current project funding and hopes for future, sorted by project *age*

GitHub provides some funding information in the form of GitHub Sponsors, but not all projects have this functionality enabled, nor is granular information about it available via the REST API.

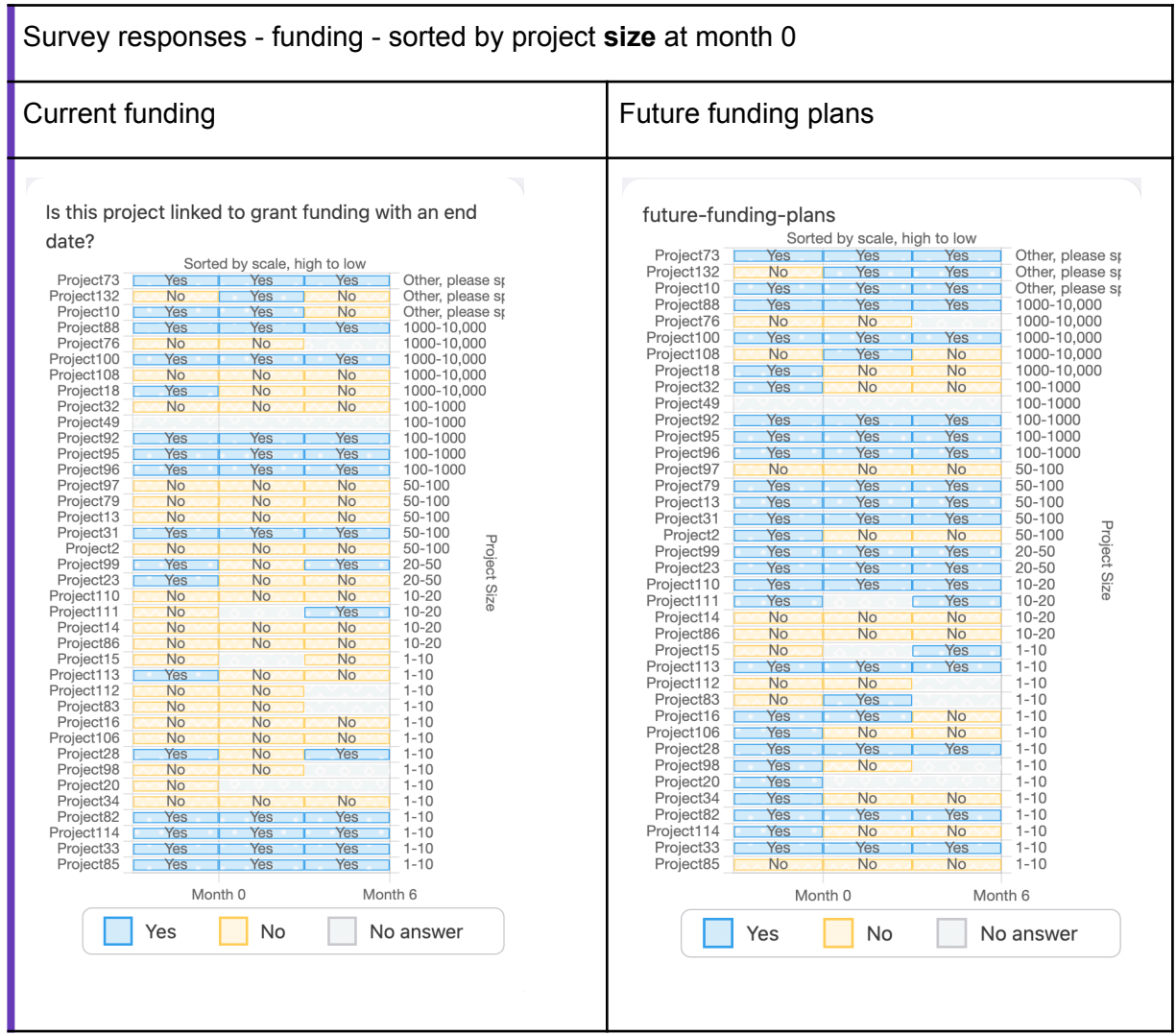

Figure 15: Current project funding and hopes for future, sorted by project *size*

17 (44% of respondents) of projects reported that they had grant funding at the start of the study, which dropped slightly to 14 (41% of respondents) at the end. This drop is <u>not</u> due to participating project leader drop-out - while month 6 and month 12 had slightly lower response rates than month 0, none of the respondents who responded "Yes" to this question in the month 0 survey failed to respond in later months.

Ten projects reported stable funding throughout the entire period, nineteen reported no funding at any stage, and 8 projects reported grant funding that ceased, stopped and re-started, or funding that started *after* the study began.

Projects that had funding at the end of the study period usually also had funding at the start of the study. Of the 38 responding projects, only two reported "No" to having grant funding in month 0, but then later reported that they had gained funding - projects 132 and 111.

One respondent observed that the grant funding question does not take into account that there may be other sources of funding available, such as industry sponsorship:

> "In our case we work collaboratively with an industry partner. In the future you may want to ask about third-party sponsorship." - Project 10

The proportion of projects intending to seek funding was much higher than the number of projects that had funding, as shown in the table below.

| | Is this project linked to grant funding with an end date? | | | Do you plan to seek funding for your project in the future? | | |
|---|---|---|---|---|---|---|
| | **# of projects** | | | **# of projects** | | |
| **Answer** | **M0** | **M06** | **M12** | **M0** | **M06** | **M12** |
| *Total* | *38 (100%)* | *34 (100%)* | *34 (100%)* | *38 (100%)* | *34 (100%)* | *34 (100%)* |
| Yes | 17 (44.74%) | 13 (38.24%) | 14 (41.18%) | 27 (71.05%) | 21 (61.76%) | 21 (61.76%) |
| No | 20 (52.63%) | 21 (61.76%) | 20 (58.82%) | 10 (26.32%) | 13 (38.24%) | 13 (38.24%) |
| No answer | 1 (2.63%) | 0 | 0 | 1 (2.63%) | 0 | 0 |

**Table 26**: Current funding and future funding plans

Project leadership: Leadership team size

> **Survey question:**
>
> How many people, including you, are in the leadership team for this project?

| Survey responses - **Leadership team size** | |
| --- | --- |
| # of people in the leadership team - sorted by project **size** | # of people in the leadership team - sorted by project **age** |

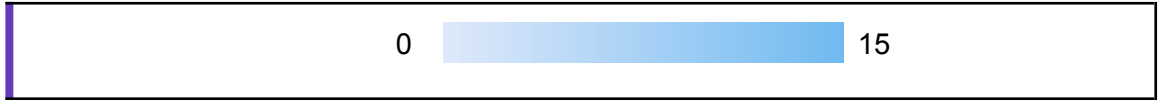

Figure 16: Number of people in the project's leadership team, sorted by project size (left) or project age (right).

| | # of projects with this number of leaders | | |
|---|---|---|---|
| **# of members in leadership team** | **M0** | **M06** | **M12** |
| *# of respondents* | *38* | *34* | *34* |
| 1 | 8 (21.05%) | 7 (18.42%) | 10 (26.32%) |
| 2 | 11 (28.95%) | 12 (31.58%) | 12 (31.58%) |
| 3 to 5 | 11 (28.95%) | 11 (28.95%) | 4 (10.53%) |
| 6 to 9 | 1 (2.63%) | 2 (5.26%) | 2 (5.26%) |
| 10+ | 2 (5.26%) | 1 (2.63%) | 3 (7.89%) |

Table 27: Leadership team sizes as reported by respondents

Around one fifth of the projects reported having one single leader, rather than a leadership team, but most frequently, a leadership team was composed of two people (reported by over one quarter of respondents - 11 or 12 projects, depending on survey timepoint). Teams of 3-5 people were not uncommon, but teams larger than this were rare. Projects 108 and 79 had the largest leadership teams, with a maximum of 12 and 15 leaders, respectively.

These results did not appear to cluster around age or project size. Project leadership: Respondent role

> **Survey questions:**
>
> **Month 0 only:** What is your role in this project? (e.g. founder, maintainer, contributor, mentor, intern)
>
> **Month 6 only:** Is your role in this project the same as it was *6 months ago*, when you filled out the previous survey?
>
> **Month 12 only:** Is your role in this project the same as it was *at the end of 2021*, when you filled out the previous survey?

When asking respondents what their role was within the project, we provided prompts for the type of answers we were looking for, such as founder, maintainer, contributor, mentor, or intern. The choice to add leading prompts was intentional - we were looking for identification of behavioural roles, rather than official job titles, which might be harder to map, e.g. "Principal software engineer" or "professor" might have meaning within an organisation, but is harder to map to day-to-day activities within an open source project. The majority of respondents identified as project founders and/or maintainers.

The counting methodology for founder, maintainer, developer, manager is inclusive of semantically similar terms: for example, the count of 22 founders shown in the table below includes all instances of the phrase "co-founder".

| Role name | # of respondents |
|---|---|
| *Respondents* | 38 |
| founder | 22 (57.89%) |
| maintainer | 21 (55.26%) |
| contributor | 2 (5.26%) |
| co-maintainer | 1 (2.63%) |
| co-PI [co-principal investigator] | 1 (2.63%) |
| co-founder | 3 (7.89%) |
| developer | 2 (5.26%) |
| co-director | 1 (2.63%) |
| co-lead | 1 (2.63%) |
| lead developer | 1 (2.63%) |
| lead maintainer | 1 (2.63%) |
| community manager | 1 (2.63%) |
| mentor | 1 (2.63%) |
| manager | 2 (5.26%) |
| PI | 1 (2.63%) |
| project lead | 1 (2.63%) |

Table 28: Project leader's self-reported roles within the project at month 0

This is a multiple choice question, so the total number of answers adds up to higher than the number of respondents.

Individual points of interest

In months 6 and 12, respondents were asked if their role was the same as last time they responded. All but one indicated that it was the same in month 6, but three others had left their original organisation by the time we ran the month 12 survey. Responses from these (erstwhile) project leaders are shown below:

| | M0 | M06 | M12 |
|---|---|---|---|
| Question | What is your role in this project? (e.g. founder, maintainer, contributor, mentor, intern) | Is your role in this project the same as it was *6 months ago*, | Is your role in this project the same as it was *at the end of 2021*, when you filled out the previous survey? |

| | | when you filled out the previous survey? | |
|---|---|---|---|
| Project 2 | maintainer,contributor | Yes | No - I am no longer the community coordinator and no longer maintain this project. |
| Project 18 | No, I left the project/company | No | Yes |
| Project 23 | co-founder,co-maintainer | Yes | Not really, funding reached an end and I am now unable to devote much time to [Project23] |
| Project 85 | founder | Yes | No |

Table 29: Responses from project leaders who reported leaving the project during the study

## Survey questions - Month 6 and 12 - midpoint and endpoint

### Changes from Month 0 - startpoint

Most questions in months 6 and 12 are identical to questions 13-37 in Month 0. Questions 1-12 from month 0 were dropped in later months, as the answers either related to the initial consent form or to questions that were not likely to change, such as "when was this project founded?"

| Question number | | Wording | | |
|---|---|---|---|---|
| M0 | M6 & M12 | M0 | M6 | M12 |
| 13 | 1 | Is this project open to contributions from the public? | Is this project *still* open to contributions from the public? | |

| 22 | 10 | What is your role in this project? (e.g. founder, maintainer, contributor, mentor, intern) | Is your role in this project the same as it was *6 months ago*, when you filled out the previous survey? | Is your role in this project the same as it was *at the end of 2021*, when you filled out the previous survey? |
|---|---|---|---|---|

Survey questions for months 6 and 12

| | Dataset variable | Question | Answer type | Answer options (if relevant) |
|---|---|---|---|---|
| 1 | project-open-contrib | Is this project still open to contributions from the public? | Yes/No/Other | |
| 2 | project-open-contrib_3_TEXT | Is this project still open to contributions from the public? | Free text for "other" option | |
| 3 | project-user-count | How many users / contributors / community members do you have at the moment? Estimate if you're not sure. Draw from analytics, GitHub, etc. as appropriate. - Selected Choice | Radio | 1-10  (1)<br>10-20  (2)<br>20-50  (3)<br>50-100  (4)<br>100-1000  (5)<br>1000-10,000  (6) |
| 4 | project-user-count_7_TEXT | How many users / contributors / community members do you have at the moment? Estimate if you're not sure. Draw from analytics, GitHub, etc. as appropriate. - Other, please specify - Text | Free text for "other" option | |
| 5 | project-user-potentl | How many users / contributors / community members would you realistically like to have in the future? Consider the size of your target audience - for example, if you're creating a community or software for a small niche interest, try to be realistic about the maximum number of people possible. - Selected Choice | Radio | 1-10  (1)<br>10-20  (2)<br>20-50  (3)<br>50-100  (4)<br>100-1000  (5)<br>1000-10,000  (6) |
| 6 | project-user-potentl_7_TEXT | How many users / contributors / community members would you realistically like to have in the future? Consider the size of your target audience - for example, if you're creating a community or software for a small niche interest, try to be realistic about the maximum number of people possible. - Other, please specify - Text | Free text for "other" option | |
| 7 | project-interns | Have you ever had paid interns working on this project, e.g. through Google Summer of Code, Outreachy, or some other internship scheme? How many? Please enter 0 if you have had no interns, or make your best guess if you are not sure. If any interns are working right now, please include them in the total. | Text | |
| 8 | you-github-user | What is your GitHub (or other code repo) user id?<br><br>We need this in order to correlate your responses with your activity on | Text | |
| | | GitHub. | | |
| 9 | you-email | What is your preferred contact email address?<br><br>We will not share this with others - we only ask so we can follow up in the future with our 6- and 12- month follow up surveys. | Text | |
| 10 | you-role | Is your role in this project the same as it was (6 months ago // at the end of 2021), when you filled out the previous survey? | Text | |
| 11 | leadership-team-size | How many people, including you, are in the leadership team for this project? | Text | |
| 12 | you-paid | Is this project part of your job? Are you paid to work on this project? - Selected Choice | Multiple options permitted | Yes, as a student  (1)<br>Yes, as a staff member with a permanent contract  (2)<br>Yes, as a staff member with a temporary or fixed-term contract  (3)<br>No  (4) |
| 13 | you-paid_5_TEXT | Is this project part of your job? Are you paid to work on this project? - Other, please specify - Text | Free text for "other" option | |
| 14 | funds-grant-funds | Is this project linked to grant funding with an end date? | Boolean Yes/No | |
| 15 | funds-others-pick-up | If you leave your role, would you expect others to pick this project up? | Boolean Yes/No | |
| 16 | funds-others- | How many others are currently paid to work on this project? This does not | Text | |
| | now | include interns. | | |
| 17 | funds-others-in-past | Have others been paid to work on the project in the past? If so, how many? This does not include interns. | Text | |
| 18 | future-funding-plans | Do you plan to seek funding for your project in the future? | Boolean Yes/No | |
| 19 | future-one-year | In an ideal situation, where do you see your project one year from now? Assume you have the funding and/or time you wish to spend on this project. - Selected Choice | Radio | still active and being maintained/updated, me still contributing  (1) still active and being maintained/updated by my community (2) still active and being maintained/updated by my colleagues  (3) finalised with occasional updates (4) wrapped up and no longer active (5) |
| 20 | future-one-year_6_TEXT | In an ideal situation, where do you see your project one year from now? Assume you have the funding and/or time you wish to spend on this project. - Other, please specify - Text | Free text for "other" option | |
| 21 | future-five-years | In an ideal situation, where do you see your project five years from now? Assume you have the funding and/or time you wish to spend on this project. - Selected Choice | Radio | still active and being maintained/updated, me still contributing  (1) still active and being maintained/updated by my community (2) still active and being maintained/updated by my colleagues  (3) |
| | | | | finalised with occasional updates (4)<br>wrapped up and no longer active (5) |
| 22 | future -five -years_6_ TEXT | In an ideal situation, where do you see your project five years from now? Assume you have the funding and/or time you wish to spend on this project. - Other, please specify - Text | Free text for "other" option | |
| 23 | future -cant -maintain | If you find yourself unable to maintain the project any longer, what do you foresee happening? Select as many as apply to your situation - Selected Choice | Multiple options permitted | My community members would keep this project running.  (1)<br>My colleagues/employees would continue to work on this  (2)<br>I would continue to provide updates in my free time  (3)<br>I would close the project down  (4)<br>I would provide periodic but rare updates when I could.  (5) |
| 24 | future -cant -maintain_ 6_TEXT | If you find yourself unable to maintain the project any longer, what do you foresee happening? Select as many as apply to your situation - Other, please specify - Text | Free text for "other" option | |
| 25 | anything -else | Final question: Anything you'd like to add that we didn't ask about? | Text | |

## Research Notice

Participants were asked to post this notice prominently on their repository README.

**Research notice**

Please note that this repository is participating in a study into sustainability of open source projects. Data will be gathered about this repository for approximately the next 12 months, starting from [insert date here].

Data collected will include number of contributors, number of PRs, time taken to close/merge these PRs, and issues closed.

For more information, please visit [the informational page](#) or download the [participant information sheet](#).

## Terminology definitions and disambiguation

This section briefly defines and disambiguates common terms used throughout this article.

| | Term | Definition |
|---|---|---|
| 1 | Code | The studies reported in this article use both qualitative and quantitative methods, which results in the linguistic overlap of the term "code", which can be used to refer to computer code or qualitative code. |
| 2 | Code - qualitative | A descriptive word, phrase, tag, or category that identifies a piece of data. In the case of these studies these data are usually interview transcripts (studies 1,2) or survey responses (studies 1,3). |
| 3 | Code - computer | The set of structured written instructions that comprises a script, program or software, often written by software engineers or programmers.<br><br>Synonyms: Our definition of computer code is intentionally broad. Some participants in study 1 did not self-identify as programmers or computer coders, |
| | | despite the fact that their survey answers asserted that they regularly wrote computer code (such as R [92], a statistical analysis package and programming language). As such, in addition to "computer code", we use the terms "**software**", "**script**", and "**program**" interchangeably to refer to the same thing. |
| 4 | Job titles: Programmer, developer, data scientist, etc. | As noted in the previous definition, computer code is written in research by many people, for many reasons. Someone might define themselves as researchers who write computer code; software developers in research; statisticians; data scientists; research software engineers; or something not included in this list.<br><br>Here, we generally use the terms "computer coder", "programmer", " (research) software engineer", or RSE for a person who writes computer code. |
| 5 | Research software | Drawing on the definition above for computer code and software, research software is any specialised computer code that is used for research purposes. This could mean generalised software that is used within day-to-day research data analysis (such as RStudio IDE[93] or Jupyter notebooks[94,95]), or it might be custom scripts or tools that have been purpose-built for a specific research task, such as the InterMine biological data warehouse[96], or the computational scripts used to create figures in this article. |
| 6 | Free software, Libre software, Open source | Throughout this article, these terms are used interchangeably for computer code that is shared freely for others to modify and re-use. Most often, we use |

| | software, FLOSS, F/LOSS, OSS, FOSS | "open source". The nuances of each of these terms is explored more deeply in the literature section. |
|---|---|---|